\documentclass[reprint,preprintnumbers,amsmath,amssymb,nofootinbib,aps,prx]{revtex4-1}

\usepackage{slashed}
\usepackage{amsmath}
\usepackage{amssymb}
\usepackage{amsthm}
\usepackage{hyperref}
\usepackage{indentfirst}
\usepackage{psfrag}
\usepackage{graphicx}
\usepackage{cleveref}
\usepackage[utf8]{inputenc}

\hypersetup{colorlinks=true, linkcolor=blue, urlcolor=blue, citecolor=blue, linktocpage=true}

\usepackage{xifthen}% Provides \isempty test
\newcommand*\diff{\mathrm{d}} % Straight differential
\newcommand*\ldiff[2][]{ \ifthenelse{\isempty{#1}}{ \diff #2}{\diff^#1#2} \,} % Differential with measure; the mandatory argument is the name of the measure, the option one is the dimension
\let\limitint\int % Only when I provide explicit limits for the integration, I need to do the spacing myself
\renewcommand{\int}{\limitint \!} % The standard integral should have correct spacing

\def\e{{\rm e}}
\def\d{\partial}

\renewcommand{\Im}{\mathop{\rm Im}\nolimits}
\renewcommand{\Re}{\mathop{\rm Re}\nolimits}

\newcommand\vf{\varphi}
\newcommand\M{\mathcal{M}}
\renewcommand\L{\mathcal{L}}
\newcommand\E{\mathcal{E}}
\newcommand\A{\mathcal{A}}

\begin{document}

\vspace{-3.0cm}
\begin{flushright}
{\small FTPI-MINN-23-07,  UMN-TH-4213/23} 
\end{flushright}
\vspace{0.5cm}

\title{Probing a Local Dark Matter Halo with Neutrino Oscillations }

\author{Tony Gherghetta$^1$}
\email{tgher@umn.edu}

\author{Andrey Shkerin$^{1,2}$}
\email{ashkerin@umn.edu}

\affiliation{$^1$School of Physics and Astronomy, University of Minnesota, Minneapolis, Minnesota 55455, USA\\$^2$William I. Fine Theoretical Physics Institute, School of Physics and Astronomy, \\ University of Minnesota, Minneapolis, Minnesota 55455, USA }

\begin{abstract}

Dark matter particles can form halos gravitationally bound to massive astrophysical objects.
The Earth could have such a halo where depending on the particle mass, the halo either extends beyond the surface or is confined to the Earth's interior. 
We consider the possibility that if dark matter particles are coupled to neutrinos, then neutrino oscillations can be used to probe an Earth dark matter halo.
In particular, atmospheric neutrinos traversing the Earth can be sensitive to a small size, interior halo, inaccessible by other means. 
Depending on the halo mass and neutrino energy, constraints on the dark matter-neutrino couplings are obtained from the halo corrections to the neutrino oscillations.

\end{abstract}

\maketitle

\section{Introduction}
\label{Intro}

Ultra-light scalar or vector bosons are well-motivated dark matter candidates. These include the QCD axion (which simultaneously addresses the strong CP problem) and a massive U(1) gauge boson or ``dark photon" \cite{ParticleDataGroup:2022pth}.
They can form a coherently oscillating background that accounts for the observed dark matter abundance \cite{Hu:2000ke,Nelson:2011sf,Hui:2016ltb,Nakayama:2019rhg}.
The interaction of Standard Model states with this background can lead to a plethora of new phenomena, including the time variation of the fundamental constants of Nature.
This opens up possibilities for dark matter searches such as, for example, in atomic clock experiments (see \cite{Safronova:2017xyt} for a review).

Furthermore, it is possible for ultra-light particles to form compact objects bound either by self-gravity and self-interaction 
\cite{Kolb:1993zz,Schive:2014dra,Hui:2016ltb,Levkov:2018kau,Veltmaat:2018dfz,Bar-Or:2018pxz,Vaquero:2018tib,Eggemeier:2019jsu},
or by the background potential of some astrophysical object \cite{Banerjee:2019epw,Banerjee:2019xuy}.
The dark matter density in such objects can be many orders of magnitude larger than the average density, thus enhancing the effects that arise from coupling to Standard Model fields.
In particular, terrestrial experiments may have increased sensitivity due to the presence of such a local halo, bound to the Earth or the Sun \cite{Banerjee:2019epw,Grote:2019uvn,Banerjee:2019xuy,Vermeulen:2021epa,Tretiak:2022ndx,Afach:2023cdc}.

Once the halo is formed with a particular mass, its stability can be maintained by the gravitational attraction of the host object. 
However, due to the complicated history of formation, the halo mass cannot be easily determined and therefore, for simplicity, we treat it as a free parameter, subject to experimental constraints.
The halo size, on the other hand, is determined by the particle mass, $m$ and the parameters of the host object.
For the Earth, a halo composed of particles with $m\lesssim 10^{-9}$ eV extends beyond the Earth's radius, thus enabling experiments conducted on the surface and in near orbit to probe it.
Instead, for $m\gg 10^{-9}$ eV, the halo size is much smaller than the Earth's radius, and therefore becomes much more difficult to detect~\cite{Banerjee:2019epw}.

In this paper, we study the possibility of using neutrinos to probe a local dark matter halo that is either larger (``big halo'') or smaller (``small halo'') than the Earth. 
Neutrino couplings to ultra-light dark matter generally result in time-varying corrections to the neutrino masses and mixing angles.
They have been studied in the literature in various terrestrial, astrophysical and cosmological setups \cite{Berlin:2016woy,Krnjaic:2017zlz,Brdar:2017kbt,Capozzi:2018bps,Liao:2018byh,Huang:2018cwo,Pandey:2018wvh,Cline:2019seo,Dev:2020kgz,Karmakar:2020yzn,Losada:2021bxx,Dev:2022bae,Salla:2022dxc,Tsai:2022jnv,Brzeminski:2022rkf,Huang:2022wmz,Alonso-Alvarez:2023tii}.
Instead, we will show that the corrections can be greatly enhanced by the overdensity of dark matter in the local halo, to the point when they are no longer small. 
The non-observation of these effects implies an upper bound on the dark matter-neutrino coupling, which becomes more stringent for heavier halos.
Furthermore, for an Earth halo with $m\gtrsim 10^{-10}$ eV, the field comprising the halo oscillates too rapidly for experiments to resolve a periodic modulation in the neutrino parameters.
Rather, we will obtain constraints on the dark matter-neutrino couplings due to the halo, which do not rely on the non-observation of such a modulation.

The neutrino is a natural, though challenging, way to measure the internal structure of the Earth (see, e.g., \cite{Akhmedov:2006hb,Agarwalla:2012uj, Winter:2015zwx,Rott:2015kwa,Bourret:2017tkw,Bakhti:2020tcj,Denton:2021rgt}), and, in particular, to detect its possible dark matter halo.
For example, an interior dark matter halo can change the average survival probability of atmospheric neutrinos passing through it.
Thus, with sufficient precision, one can use neutrinos coming from different directions to scan the spatial profile of the halo. 
Depending on the halo mass and the strength of the neutrino-halo interactions, the inner halo region can cause nonadiabatic neutrino oscillations.
These nonadiabatic effects are due to the rapid time variation of the field comprising the halo; interestingly, they can both enhance and suppress distortions to the vacuum neutrino oscillations.
We will also see that at sufficiently large neutrino energies, the magnitude of the deviation from the vacuum oscillation probability becomes energy independent.

We will focus on two-flavour oscillations to compute these effects for both a big and small halo surrounding the Earth.
For simplicity, we will neglect the Standard Model neutrino-matter interactions.
We will see that, unlike the Mikheyev-Smirnov-Wolfenstein (MSW) resonances due to the matter potential of the Earth, the nonadiabatic correction to the oscillation probability due to the dark matter halo can be large in a broad range of neutrino energies.

In Section~\ref{sec:halo}, we revisit the properties of a nonrelativistic halo, gravitationally bound by an external body where we will remain agnostic about the particular halo formation mechanism.
The halo is assumed to comprise massive (pseudo)-scalar particles and no self-interaction or couplings to Standard Model fields (except to the neutrino) will be needed to discuss the halo effects.
In Section~\ref{sec:neutrino}, we calculate the corrections to neutrino oscillations in the presence of the halo.
We consider the cases when the scalar contributes to the neutrino mass via a dimension-four Yukawa coupling, or to the neutrino momentum via a dimension-five derivative coupling. These couplings naturally arise in DFSZ-type axion models as well as in flavor-axion models (see, e.g., \cite{Mohapatra:1982tc, Kelly:2018tyg,Cox:2021lii}). Next, in Section~\ref{sec:Pheno}, we focus on the Earth as the host body and  study the survival probability of neutrinos propagating in the time- and space-varying halo profile. We study both the small corrections within perturbation theory and the adiabatic approximation, as well as nonadiabatic resonance effects.

In Section~\ref{sec:vector} we consider a halo made of massive real vector particles, such as dark photons. We solve the equations of motion and obtain vector field configurations describing a nonrelativistic, radially-polarised halo, gravitationally bound to the host body.
Under the assumption that the vector field couples to the neutrino current, we repeat the neutrino oscillation analysis in the background of the vector halo and obtain constraints on the coupling parameters. Our concluding remarks are given in Section~\ref{sec:concl}.

\section{Local Scalar Halo}
\label{sec:halo}

\begin{figure}[t]
\center{
\includegraphics[width=0.95\linewidth]{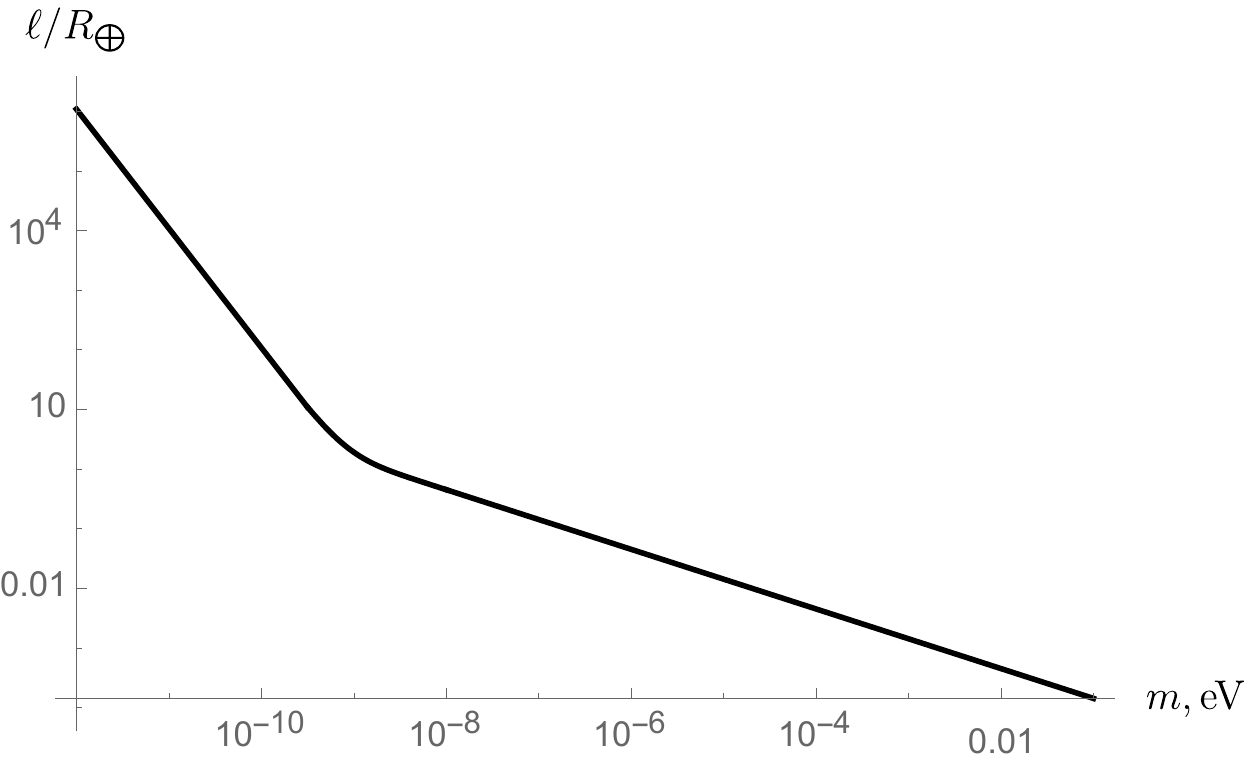}
\caption{The size of the Earth's dark matter halo (in units of $R_\oplus$) as a function of the scalar field mass, $m$.}
\label{fig:l1}}
\end{figure}

Consider a free, massive, real scalar field $\vf$ in the gravitational background generated by the host astrophysical object. In the nonrelativistic, weak-field regime, the line element, assuming spherical symmetry, takes the form 
\begin{equation}\label{ds}
\begin{aligned}
    & \diff s^2=-N(r) c^2\diff t^2 + \frac{\diff r^2}{N(r)} + r^2\diff\Omega^2 \;, \\
    & N(r)=1+\frac{2\Phi(r)}{c^2} \;,
    \end{aligned}
\end{equation}
where $\Phi(r)$ is the Newtonian gravitational potential, 
and $c$ is the speed of light.
The real scalar field $\vf$ with mass $m$ can be decomposed into the nonrelativistic wavefunction $\Psi$ as~\footnote{Note that $\Psi(r,t)$ represents a classical solitonic configuration (halo).
This is ensured by the large occupation number of the field modes comprising the halo. }
\begin{equation}\label{phi_nonrel}
    \vf(r,t)=\sqrt{\frac{2c}{m}}\left(\Psi(r,t) \e^{-imc^2t} + c.c.\right) \;.
\end{equation}
To leading order in $c$, the equation of motion for $\vf$ in the background (\ref{ds}) becomes
\begin{equation}\label{eom1}
    i\dot{\Psi}(r,t)=-\frac{1}{2m}\Delta \Psi(r,t) 
    +m\,\Phi(r)\Psi(r,t) \; 
\end{equation}
where the dot denotes the time derivative.
We are interested in stationary, bound-state, spherically-symmetric solutions of this equation, $\Psi(r,t)=\psi(r)\e^{-i E t}$ where $|E|\ll mc^2$ is the nonrelativistic energy.
The host astrophysical object is assumed to be a sphere of constant density with mass $M$ and radius $R$.
Introducing the dimensionless variables
\begin{equation}\label{units}
    x=r/R \;, ~~~ \M=Gm^2MR \;, ~~~  \E=EmR^2 \;,
\end{equation}
where $G$ is Newton's constant, \cref{eom1} can then be conveniently written as
\begin{equation}\label{eom2}
    -\frac{1}{x^2}\frac{\diff }{\diff x}\left(x^2\frac{\diff\psi}{\diff x}\right)+2(\M\Tilde{\Phi}-\E)\psi =0 \;,
\end{equation}
where
\begin{equation}\label{PhiTilde}
    \tilde{\Phi}(x)=\left\lbrace \begin{array}{ll}
         \frac{1}{2}(x^2-3)  & ~~ x<1 \\
        -\frac{1}{x}  & ~~x>1
    \end{array} \right. \;.
\end{equation}
The solution $\psi(x)$ to the equation of motion \eqref{eom2} is assumed to be regular at $x=0$ and to vanish as $x\to\infty$. 
Furthermore, requiring that the solution $\psi(x)$ is smoothly continuous at $x=1$ leads to a discrete set of allowed bound-state energies $\E_n, n=0,1,2\dots$. The solution can be written in terms of the confluent hypergeometric functions:
\begin{equation}
\label{sol1}
\psi_n\propto
\left\lbrace\begin{array}{ll}
    \e^{-\frac{1}{2}\sqrt{\mathcal{M}}x^2}\times \\
    \quad{}_1 F_1\left(\frac{3}{4}(1-\sqrt{\mathcal{M}})+\frac{|\mathcal{E}_n|}{2\sqrt{\mathcal{M}}},\frac{3}{2},\sqrt{\mathcal{M}}x^2 \right) & x<1\;,\\
    %\label{sol1a} \\
     \e^{-\sqrt{2|\mathcal{E}_n|}x}\,U\left(1-\frac{\mathcal{M}}{\sqrt{2|\mathcal{E}_n|}} , 2, 2\sqrt{2|\mathcal{E}_n|}x\right) &x>1\;. 
     %\label{sol1b}
\end{array}\right.
\end{equation}
For our purposes, the ground state $\psi_0$ with the lowest energy $\E_0$ will be the most relevant solution. 
The physical halo size can be defined as $\ell\equiv Rx_\ell$, where $x_\ell$ is the distance at which the amplitude of the profile (\ref{sol1}) decreases by a factor of $1/\e$.
From the asymptotic behavior of the functions in the solution (\ref{sol1}) we then find
\begin{equation}\label{x_st}
    |\mathcal{E}_0|\simeq\left\lbrace \begin{array}{l}
        \frac{1}{2}\M^2 \\
        \sqrt{2}\M
    \end{array} \right. \;,~~~
    x_\ell\simeq\left\lbrace \begin{array}{ll}
        \frac{1}{\mathcal{M}}  & ~~\M\ll 1  \\
        \frac{\sqrt{2}}{\M^{1/4}} & ~~\M\gg 1 
    \end{array} \right. \;.
\end{equation}
In what follows, we will focus on the Earth (with mass $M_{\oplus}$, radius $R_{\oplus}$) as the host of the dark matter halo. From \cref{units} we obtain 
\begin{equation}\label{M_gen}
    \M\simeq \left(\frac{m}{10^{-9}\:\text{eV}}\right)^2\left(\frac{MR}{M_{\oplus}R_{\oplus}}\right) \;.
\end{equation}
Thus, $\M\sim 1$ corresponds to a nano-eV mass, dark matter particle with a physical halo size $\ell\sim R_{\oplus}$.
From \cref{x_st} it also follows that
\begin{subequations}\label{l_m}
\begin{align}
    &\ell\sim R_{\oplus} \left( \frac{10^{-9}\:\text{eV}}{m}\right)^2  ~~~~~ m\ll 10^{-9}\:\text{eV} \;, \label{l_m_1}\\
    & \ell\sim R_{\oplus} \left( \frac{10^{-9}\:\text{eV}}{m}\right)^{1/2}  ~~~ m\gg 10^{-9}\:\text{eV} \;. \label{l_m_2}
\end{align}
\end{subequations}
This behaviour is illustrated in Fig.~\ref{fig:l1}, which agrees with the more qualitative analysis of Ref.~\cite{Banerjee:2019epw}. Note that the analytic solution for the halo profile (\ref{sol1}) does not apply for a more realistic distribution of matter in the Earth \cite{DZIEWONSKI1981297}; however, the parametric dependence in \cref{M_gen,l_m} remains valid.

When $m\lesssim 10^{-10}$ eV, the halo extends much beyond the Earth's radius.
In this case, local experiments are not sensitive to the spatial profile of the halo, and can only probe the amplitude $f_0$ of the time variations of the field $\vf$ in the halo center.
In this paper, we consider a ``big'' local halo that extends not far from the Earth's surface, corresponding to a mass $m\sim 10^{-10}$\,eV.
This is because for much smaller masses, the larger size halo (using \eqref{l_m_1}) would likely be disrupted by the gravitational pull of the Sun.
Note that the analysis below regarding the neutrino propagating in a constant-amplitude background is readily applicable to the case when the local halo inhabits the Solar System and is hosted by the Sun, corresponding to $m\lesssim 10^{-13}$ eV.
The results of the previous studies can also be adapted to this case \cite{Krnjaic:2017zlz,Capozzi:2018bps,Dev:2020kgz,Losada:2021bxx}.
Nevertheless, our main interest is to explore the sensitivity of terrestrial experiments to the spatial halo profile, rather than the background value, for which $m\gtrsim 10^{-9}$ eV is an appropriate mass range.
Interestingly, this includes the QCD axion mass range $10^{-12}{\rm eV}\lesssim m \lesssim 10^{-3}{\rm eV}$ corresponding to axion decay constants $10^9{\rm GeV}\lesssim f_a \lesssim 10^{18}{\rm GeV}$.

The halo mass is estimated as $M_{\text{halo}}\sim \ell^3m^2f_0^2$, where $f_0^2$ is proportional to the occupation number of the field modes comprising the halo and can be very large.
If $m\sim 10^{-10}$\,eV, most of the halo is located inside the moon's orbit, and the constraint on $M_{\text{halo}}$ (and, hence, on $f_0$) arises from lunar laser ranging, $M_{\text{halo}}\lesssim 10^{16}$ kg \cite{Adler:2008rq}.
For small halos $\ell \lesssim R_\oplus\, (m\gtrsim 10^{-9}$~eV), there is no such constraint, and we will consider $M_{\rm halo}\leqslant 0.1M_{\oplus}$ to ensure that the halo contributes negligibly to the gravitational potential of the Earth.
We will next employ neutrino interactions to probe both big and small halos.

\section{Neutrino Interaction with the Halo}
\label{sec:neutrino}

We will study the effect on the oscillations of left-handed active neutrinos $\psi_{L}$ from their interactions with the halo. 
We will consider two simple scalar-neutrino interaction terms whose effect can be qualitatively different.
The first is the dimension-four operator
\begin{equation}\label{L_int2}
\L_{4,\rm int}=-yh_{ab}\vf\bar{\psi}_{La}\psi^C_{Lb} +h.c.
\end{equation}
where $\psi^C_{La}$ denotes the charge conjugate,
$a,b$ are flavour indices, $h_{ab}$ is a complex and symmetric flavour matrix whose values are assumed to be of order one, and $y$ is a small dimensionless coupling.
Furthermore, $\vf$ is the background halo configuration (\ref{phi_nonrel}), (\ref{sol1}) which we rewrite as follows 
(switching to the natural units with $c=1$)
\begin{equation}\label{Halo}
\vf(r,t)=f(r)\cos(mt+\delta) \;,
\end{equation}
where $\delta$ is the phase of the halo at the moment $t=0$ of neutrino production.
The interaction \eqref{L_int2} modifies the dispersion relation of neutrinos by shifting the neutrino mass. Interactions of this type generically arise in DFSZ-type axion models (see, e.g., \cite{Mohapatra:1982tc, Kelly:2018tyg}). 

Another possibility is the dimension-five derivative operator\footnote{If $\vf$ is a pseudoscalar, one should insert $\gamma^5$ into the neutrino current in (\ref{L_int1}). Since we consider oscillations of ultrarelativistic active neutrinos, this does not change the subsequent analysis. }
\begin{equation}\label{L_int1}
\L_{5,\rm int}=-\frac{g_{ab}}{\Lambda_5} \,\d_\mu\vf\, \bar{\psi}_{La}\gamma^\mu\psi_{Lb} \;,
\end{equation}
where $\Lambda_5$ is the UV scale at which the interaction is generated.
The coupling matrix, $g_{ab}$ is Hermitian and we assume that its values are of order one. Interestingly, axion models, which address both the axion quality problem and the flavor hierarchies in the Standard Model, generate interactions of the type \eqref{L_int1} with $\Lambda_5\gtrsim 10^{13}$~GeV~\cite{Cox:2021lii}.
In the background \eqref{Halo}, the interaction \eqref{L_int1} modifies the neutrino dispersion relation by shifting the neutrino momentum. 
In general, both matrices $h$, $g$ do not commute with the neutrino mass matrix in the flavor basis, $m_\nu$.

We adopt the plane-wave treatment of neutrino oscillations and neglect the effects of neutrino dispersion and decoherence.
The evolution equation for the ultrarelativistic neutrino wavefunction $\nu_a$ in the flavour basis then takes the form
\begin{equation}\label{EvolEq}
i\frac{\diff\nu_a}{\diff z}=H_{ab}\nu_b \;.
\end{equation}
Here $z$ denotes the direction of neutrino propagation, and the Hamiltonian $H$ is given by
\begin{equation}\label{H0}
    H=\frac{1}{2E}U_0\text{diag}(0,\Delta m_{0,21}^{2},...)U_0^{\dagger} + \Delta H \;,
\end{equation}
where $E$ is the mean neutrino energy, $U_0$ is the vacuum neutrino mixing matrix, $\Delta m^2_{0,ij}$ are the mass-squared differences, and $\Delta H$ depends on the choice of the scalar-neutrino interaction.
In particular, the interaction (\ref{L_int2}) results in the following contribution to the Hamiltonian:
\begin{equation}\label{DeltaH2}
    \Delta H_4=\frac{y}{E}\vf (h^\dagger m_\nu + m_\nu^\dagger h) +\frac{2y^2}{E}\vf^2 h^\dagger h \;,
\end{equation}
where $\vf$ is given in \cref{Halo}.

Next, consider the interaction (\ref{L_int1}).
In the scalar halo background hosted by the Earth, the temporal component of the current in \cref{L_int1} dominates over the gradient part.
Indeed, using \cref{Halo} one can estimate $|\d_0\vf|\sim mf_0$, $|\nabla\vf|\sim f_0/\ell$, and from \cref{l_m} we obtain $m\ell\gtrsim 10^4$.
Hence, the interaction (\ref{L_int1}) is analogous to the MSW effect; it contributes to the Hamiltonian as follows:
\begin{equation}\label{DeltaH1}
    \Delta H_5=\frac{m}{\Lambda_5}g\vf  \;,
\end{equation}
where $\dot \varphi=m\varphi$ upon shifting the phase.
The Hamiltonian \eqref{H0} can be written as
\begin{equation}\label{H1}
    H=\frac{1}{2E} U\text{diag}(0,\Delta m_{21}^2,...)U^\dagger \;,
\end{equation}
where the unitary matrix $U$ diagonalizes the full Hamiltonian, and $\Delta m^2_{ij}$ are the corresponding $z$-dependent eigenvalues that have been shifted by the scalar field background.

Within the adiabatic approximation, the oscillation probability at the baseline $L$ is given by
\begin{equation}\label{P_osc}
    P_{ab}(L)=\left\vert \sum_i U_{a i}(0)\e^{-\frac{i}{2E}\limitint_0^L\diff z\:m_i^2(z)}U^*_{b i}(L)\right\vert^2 \;,
\end{equation}
where $z=0,L$ are the neutrino production and detection locations, respectively.
For simplicity, we will focus on two-flavour oscillations.
The survival probability for flavor $a$ is then
\begin{equation}\label{P_surv}
\begin{aligned}
&P_{aa}(L)=\frac{1}{2}\biggl\lbrace 1+\cos2\theta(0)\cos2\theta(L) \\
&\qquad\qquad\qquad+\sin2\theta(0)\sin2\theta(L)\cos\left( 2X_{\text{eff}} \right) \biggr\rbrace \;,
\end{aligned}
\end{equation}
where $X_{\text{eff}}=\Delta m_{\text{eff}}^2L/(4E)$ and the effective mass-squared difference is
\begin{equation}\label{DeltaMEff}
\Delta m_{\text{eff}}^2=\frac{1}{L}\limitint_0^L\diff z\,\Delta m^2(z) \;.
\end{equation}
In the adiabatic regime, it is useful to expand the oscillation parameters in \eqref{P_osc}-\eqref{DeltaMEff} around their vacuum values. 
To analyse both types of scalar-neutrino interactions \eqref{L_int2} and \eqref{L_int1}, we define
\begin{equation}\label{Beta}
    \beta_4\equiv\frac{y \sum m_\nu }{2E} \;, ~~~ \beta_5\equiv\frac{m}{2\Lambda_5} \;,
\end{equation}
where $\sum m_\nu$ is the sum of the physical active neutrino masses.
Using $\beta\ll 1$ as the perturbative parameter (where $\beta$ denotes either $\beta_4$ or $\beta_5$), the first three terms in the expansion of the mixing angle and the mass-squared difference are:
\begin{subequations}\label{ParamExp}
\begin{align}
    &\theta(z)=\theta_{0}+\beta \theta_{1}(z)+\beta^2 \theta_{2}(z) +... \label{ParamExp1} \\
    &\Delta m^2(z)=\Delta m_{0}^{2}+\beta \Delta m_{1}^{2}(z)+\beta^2\Delta m_{2}^{2}(z)+... \label{ParamExp2}
\end{align}
\end{subequations}
Substituting into \cref{H1}, it is straightforward to obtain
\begin{subequations}\label{corr}
\begin{align}
&  \theta_1(z)=\frac{\Gamma(z)}{2\Delta m_0^{2}}\A_{\theta 1} \;, ~~~~~~~\Delta m^{2}_1(z)=\Gamma(z)\A_{m 1} \;, \label{corr1} \\
& \theta_2(z)=\frac{\Gamma^2(z)}{\left(2\Delta m_0^{2}\right)^2}\A_{\theta 2} \;, 
~ \Delta m_2^{2}(z) = \frac{\Gamma^2(z)}{2\Delta m_0^{2}}\A_{m2} \;. \label{corr2}
\end{align}
\end{subequations}
The function $\Gamma(z)\equiv 4E\vf(r(z),t(z))$ depends on the time- and space-varying halo profile probed by the neutrino.
For ultrarelativistic neutrinos, $t=z$, and using \cref{Halo} we obtain
\begin{equation}\label{Gamma}
\Gamma(z)= 4Ef(z)\cos(mz+\delta ) \;.
\end{equation}
In \cref{corr}, the dimensionless parameters $\A_{\theta,m}$ are of order one.
They depend on the vacuum mixing angle and the matrix elements, $h_{ab}$, $m_{\nu ab}/\sum m_\nu$ or $g_{ab}$, for the interactions (\ref{L_int2}) or (\ref{L_int1}), respectively.
Their exact form is not important for the subsequent analysis; for completeness, we quote them in Appendix~\ref{sec:AppA}.

\section{Neutrino Propagation through the Halo}
\label{sec:Pheno}

\subsection{Adiabatic oscillations in the big halo}
\label{ssec:ad}

We start with the big halo case (corresponding to a halo size $l\gtrsim R_\oplus$) which can be probed with reactor or accelerator neutrinos.
For the benchmark value $m=10^{-10}$ eV, 
the background scalar field oscillates too rapidly for experiments to resolve any periodic modulation in the neutrino data, and therefore the oscillation probability should be averaged over the phase $\delta$ of the halo,
\begin{equation}\label{P_surv_av}
\langle P_{aa} \rangle_\delta \equiv \frac{1}{2\pi }\limitint_0^{2\pi} \diff\delta \:P_{aa} \;.
\end{equation}
We consider neutrino energies in the range 1 MeV--\,1 GeV which covers both reactor and accelerator neutrino experiments. Thus, for the vacuum neutrino mass-squared difference we take the value appropriate for these experiments, $\Delta m_0^2=2.5\times 10^{-3}$ eV$^2$ \cite{ParticleDataGroup:2022pth}.

\begin{figure}[t]
\center{
\includegraphics[width=0.85\linewidth]{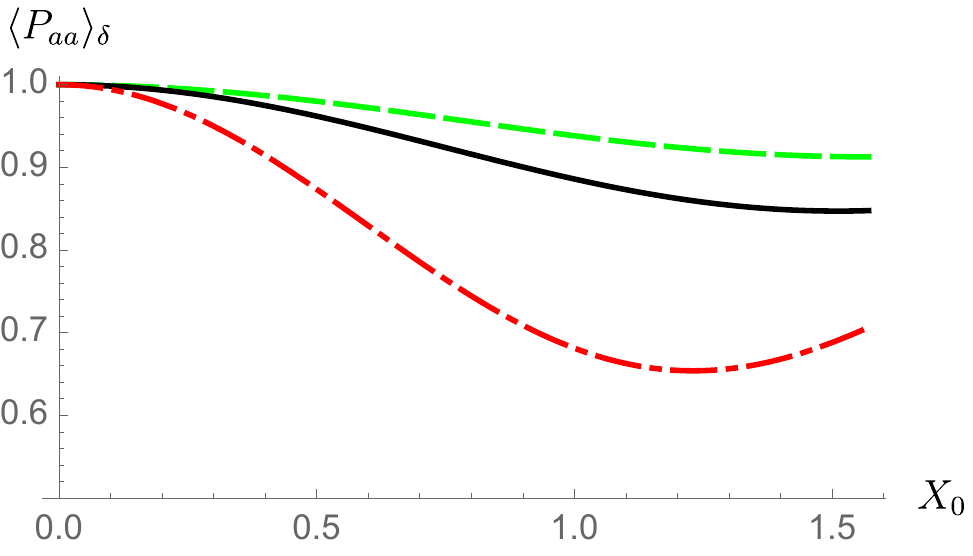}
\caption{The averaged neutrino survival probability \eqref{P_surv2} in the near-constant background, within the validity of perturbation theory and the adiabatic approximation, as a function of $X_0=\pi L/L_0^{\rm osc}$.
We assume that the neutrino couples to the halo via the derivative interaction (\ref{L_int1}) and take $\sin^22\theta_0=0.087$, $g_{11}=0.5$, $g_{12}=i$, $g_{22}=0$, $\eta=0.1$ and $\epsilon\eta=0.1$ (black solid), $0.25$ (red dot-dashed).
The green dashed line denotes the vacuum probability.
}
\label{fig:osc_pert}}
\end{figure}

First, we discuss perturbative corrections to the survival probability in the adiabatic approximation.
It is convenient to introduce the following parameters,
\begin{equation}\label{epsilon}
\epsilon \equiv \frac{\beta f_0}{m} \sim \left(\frac{\beta}{10^{-22}}\right)\left(\frac{m}{10^{-10}\:\text{eV}}\right)\left( \frac{M_{\text{halo}}}{10^{15}\:\text{kg}} \right)^{1/2} \;,
\end{equation}
\begin{equation}\label{eta}
\eta \equiv\frac{mE}{\Delta m_0^2}=\left(\frac{2.5\times 10^{-3}\:\text{eV}^2}{\Delta m_0^2}\right)\left(\frac{m}{10^{-10}\:\text{eV}}\right)\left(\frac{E}{25\:\text{MeV}}\right) ,
\end{equation}
where $M_{\text{halo}}\sim\ell^3m^2f_0^2$ and \cref{l_m_1} has been used. 
The parameter $\epsilon$ plays the role of an expansion parameter, while $\eta$ determines the number of halo oscillations in one neutrino oscillation length.
After integrating \eqref{P_surv_av} over $\delta$ using \cref{P_surv}, the linear in $\beta$ correction to the survival probability vanishes. 
To second order in $\beta$, one obtains
\begin{equation}\label{P_surv2}
\begin{aligned}
& \langle P_{aa}\rangle_\delta = 1-\sin^22\theta_0\sin^2X_0 \\
& -2 \epsilon ^2 \biggl(A_{m1}^2 \sin ^2 2 \theta _0
   \cos (2 X_0) \sin^2(2\eta  X_0)\\
&\qquad +\eta  A_{\theta 1} A_{m1} \sin 4 \theta
   _0 \sin (2 X_0) \sin (4 \eta  X_0)\\
   &\qquad+2 \eta ^2 \biggl[ X_0 A_{m2} \sin ^2 2 \theta_0 \sin (2 X_0)\\
   &\qquad+2 A_{\theta 1}^2 \bigl( \cos 4 \theta _0
   \sin^2X_0 \cos ^2(2 \eta  X_0)\\
   & \qquad+\cos^2X_0 \sin ^2(2 \eta  X_0)\bigr)
   +A_{\theta 2} \sin 4 \theta _0 \sin^2X_0\biggr]\biggr) \;,
\end{aligned}
\end{equation}
where $X_0=\pi L/L_0^{\rm osc}$ and $L_0^{\rm osc}=4\pi E/ \Delta m_0^2$ is the vacuum oscillation length. 
Note that the correction to the vacuum oscillation probability is qualitatively different depending on the asymptotic limits of $\eta$.
When $\eta\ll 1$ (halo oscillation much slower than the neutrino), the correction is simply due to the constant background potential (similar to the MSW effect).
Expanding \cref{P_surv2} for small $\eta$, the leading correction term is $\propto\epsilon^2\eta^2$, and hence the perturbation expansion is valid until $\epsilon\sim\eta^{-1}\gg 1$.
This can be seen in Fig.~\ref{fig:osc_pert}, where (\ref{P_surv2}) is plotted at $E=2.5$ MeV and several values of $\epsilon\eta$. 
Neutrinos of similar energies are studied in medium baseline reactor experiments, and thus we assume $\sin^22\theta_0=\sin^22\theta_{13}=0.087$, relevant for the survival probability $P_{ee}$ \cite{ParticleDataGroup:2022pth}.
For definiteness, in Fig.~\ref{fig:osc_pert} we use the dimension-five coupling (\ref{L_int1}).

In the second case, when $\eta\gg 1$ (halo oscillates much faster than the neutrino), the probability is modulated by small ``wiggles'' of frequency $\sim \eta X_0/L\sim m$.~\footnote{Note that the length scale of the halo oscillations, $\sim m^{-1}$, is still much larger than the effective length of the neutrino wavepacket (see, e.g., \cite{Akhmedov:2009rb,Akhmedov:2022bjs}), and does not spoil the plane-wave treatment of neutrino oscillations according to \cref{EvolEq}.} 
Expanding at large $\eta$, one again finds that the leading correction is $\propto\epsilon^2\eta^2$, which would lead to the conclusion that the perturbation expansion remains valid until $\epsilon\sim\eta^{-1}$.

However, the above analysis is applicable only as long as nonadiabatic effects are small.
The adiabatic approximation is controlled by the gradient of the instantaneous mixing angle: $\theta'(z)\ll\Delta m^2(z)/E$.
The function $\theta'(z)$ not only depends on the spatial gradient of the halo but also on its much more rapid temporal variation.
From \cref{corr1,Gamma} we find that inside the halo $\theta'(z)\sim \beta Emf_0/\Delta m_0^2$.
Hence, the expression (\ref{P_surv}) and the perturbative result (\ref{P_surv2}) are valid, provided
\begin{equation}\label{CondAd}
\epsilon\eta^2\ll 1\;.
\end{equation}
Thus, for neutrinos with $\eta\gg 1$, \cref{P_surv2} is only valid for
$\epsilon\lesssim\eta^{-2}$.
For larger halo amplitudes (or larger neutrino energies), the time variation of the oscillation probability leads to multiple resonances during the neutrino propagation and, in general, needs to be treated numerically.
We will next study this case.

\subsection{Nonadiabatic regime in the big halo}
\label{sssec:nonad}

When the analytic expression (\ref{P_surv2}) is no longer valid, either because perturbation theory or the adiabatic approximation breaks down, one has to solve the evolution equation (\ref{EvolEq}) numerically.
The survival probability $P_{aa}$ of flavour $a$, at a distance $L$ with boundary conditions $\nu_a(0)=1$, $\nu_b(0)=0$ is then given by $|\nu_a(L)|^2$.
Before presenting the numerical results, it is instructive to analytically estimate the deviation from the vacuum probability in the nonadiabatic regime for neutrino energies $E\gg 25$ MeV, corresponding to $\eta\gg 1$.
In this case, as discussed in Section~\ref{ssec:ad}, the adiabatic regime breaks down at $\epsilon\sim \eta^{-2}$, long before the correction to the vacuum oscillations becomes sizeable.
The probability behavior at larger values of $\epsilon$ is qualitatively different for the dimension-four (\ref{L_int2}) and dimension-five (\ref{L_int1}) interactions, and therefore we will treat them separately.

\begin{figure}[t]
\center{
\includegraphics[width=0.99\linewidth]{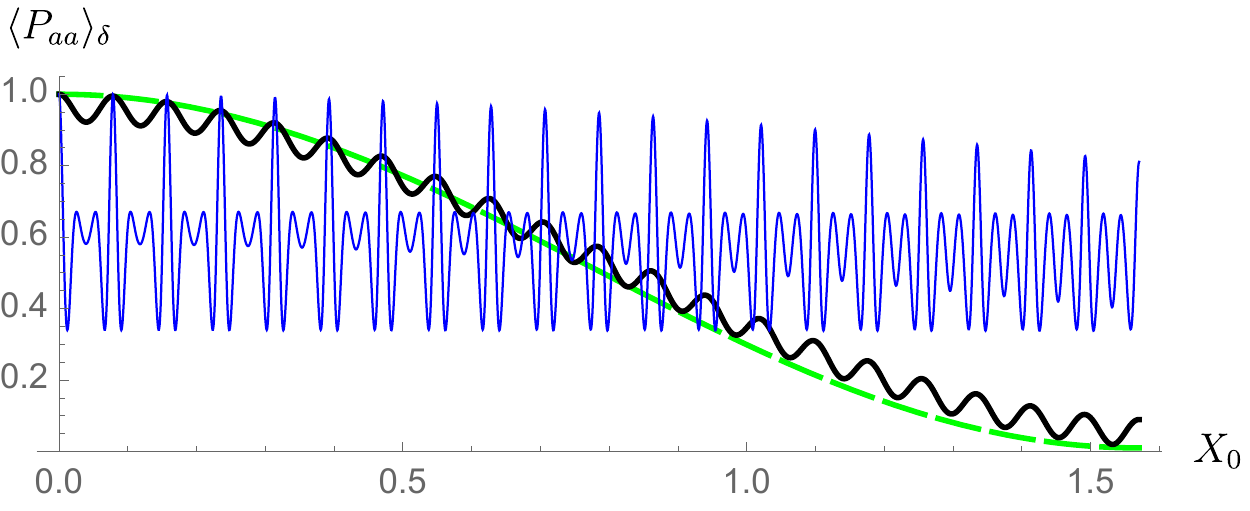}
\caption{The averaged neutrino survival probability \eqref{P_surv_av} in the oscillatory halo background, where the neutrino couples to the halo via the derivative interaction (\ref{L_int1}), assuming $\sin^22\theta_0=0.99$, $g_{11}=0.5$, $g_{12}=i$, $g_{22}=0$, $\eta=40$ and $\epsilon=0.1$ (black), $1.0$ (thin blue). 
The green dashed line denotes the vacuum probability. 
}
\label{fig:osc_nonad5}}
\end{figure}

Consider first the derivative coupling (\ref{L_int1}).
Interestingly, in this case, nonadiabatic effects tend to suppress the correction until $\epsilon\sim 1 \gg \eta^{-1}$.
To see this explicitly, we rotate to the mass basis in \cref{EvolEq} with the vacuum mixing matrix, $\nu_i=U^{\dagger}_{0,ia}\nu_a$. 
Using \cref{DeltaH1,Beta}, we obtain
\begin{equation}\label{EvolEq2}
i\frac{\diff\nu_i}{\diff z}=\frac{m_i^2}{2E}\nu_i +2\beta_5 f_0\cos (mz+\delta)\: \tilde{g}_{ij}\nu_j \;,
\end{equation}
where $\tilde g = U^\dagger_0 g U_0$.
We expand the mass eigenstates around their vacuum values,
\begin{equation}
\label{eq:masseigen}
\nu_i(z)=(1+\Delta\nu_i(z))\e^{-i\frac{m_i^2 z}{2E}} \;,
\end{equation}
assuming that $|\Delta\nu_i(z)|\ll 1$.
Substituting \cref{eq:masseigen} in \cref{EvolEq2} and assuming $\eta\gg 1$, we find that the deviation accumulated over one neutrino oscillation period is
\begin{equation}
|\Delta\nu_i(L_{0}^{\text{osc}})|\sim \epsilon|\tilde g_{ij}\sin (4\pi\eta)| \;,
\end{equation}
where, in the two-flavour scheme, $i,j=1,2$ and $i\neq j$.
Thus, for neutrinos with $E\gg 25$ MeV, the size of the correction due to the halo is controlled by the parameter $\epsilon$.
Note again that, even though at $\eta^{-2}\lesssim\epsilon \ll 1$ the correction to the vacuum oscillation is small, the neutrino propagation is governed by nonadiabatic effects.
The neutrino experiences two resonances at every cycle of the halo time variation; however, their combined effect is small unless $\epsilon\gtrsim 1$.
This behavior is illustrated in Fig.~\ref{fig:osc_nonad5}, which shows the numerical solution for the survival probability, averaged over the halo phase, at $E=1$ GeV (corresponding to $\eta=40$) and several values of $\epsilon$.
The neutrinos with these energies are typical in long baseline accelerator experiments, and hence for the mixing angle we adopt the value $\sin^22\theta_0=\sin^22\theta_{23}=0.99$, relevant for the survival probability $P_{\mu\mu}$ \cite{ParticleDataGroup:2022pth}.
From Fig.~\ref{fig:osc_nonad5} we see that the halo time variation induces secondary oscillations in the vacuum neutrino oscillations.
In the limit $\epsilon\gg 1$, the probability, which is averaged over these secondary oscillations, tends to $1/2$.

\begin{figure}[t]
\center{
\includegraphics[width=0.99\linewidth]{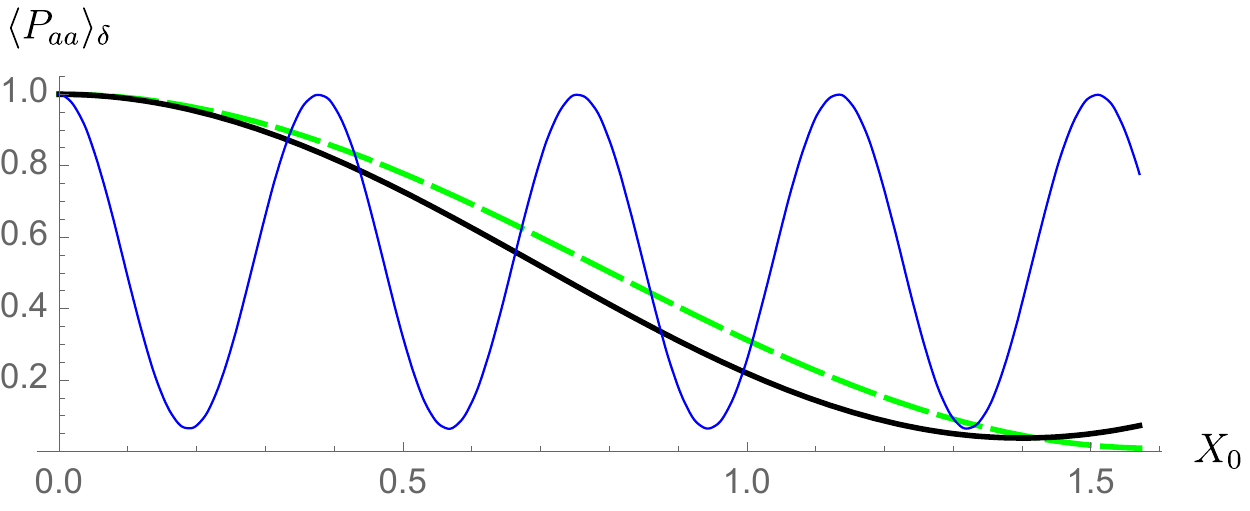}
\caption{The averaged  neutrino survival probability \eqref{P_surv_av} in the oscillatory halo background, where the neutrino couples to the halo via the interaction (\ref{L_int2}), assuming $\sin^22\theta_0=0.99$, $\Delta m_0^2=2.5\times 10^{-3}$ eV$^2$, $\sum m_\nu=0.1$ eV, $h_{11}=0.5$, $h_{12}=i$, $h_{22}=0$, $\eta=40$ and $\epsilon\eta=0.5$ (black), $2.0$ (thin blue). 
The green dashed line denotes the vacuum probability. 
}
\label{fig:osc_nonad4}}
\end{figure}

\begin{figure*}[t]
    \center{
        \begin{minipage}[h]{0.45\linewidth}
            \center{\includegraphics[width=0.99\linewidth]{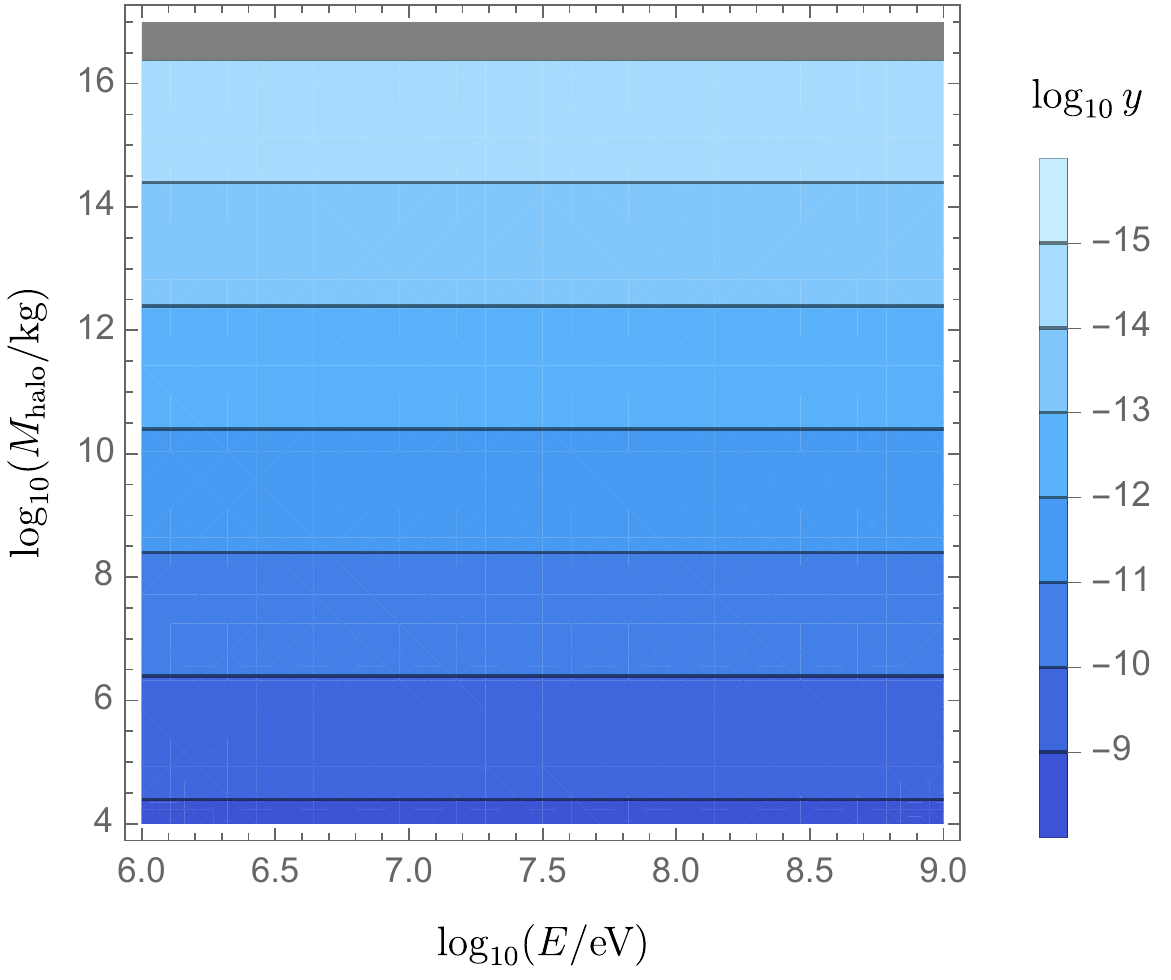}}
        \end{minipage}
		\hfill
		\begin{minipage}[h]{0.45\linewidth}
			\center{\includegraphics[width=0.99\linewidth]{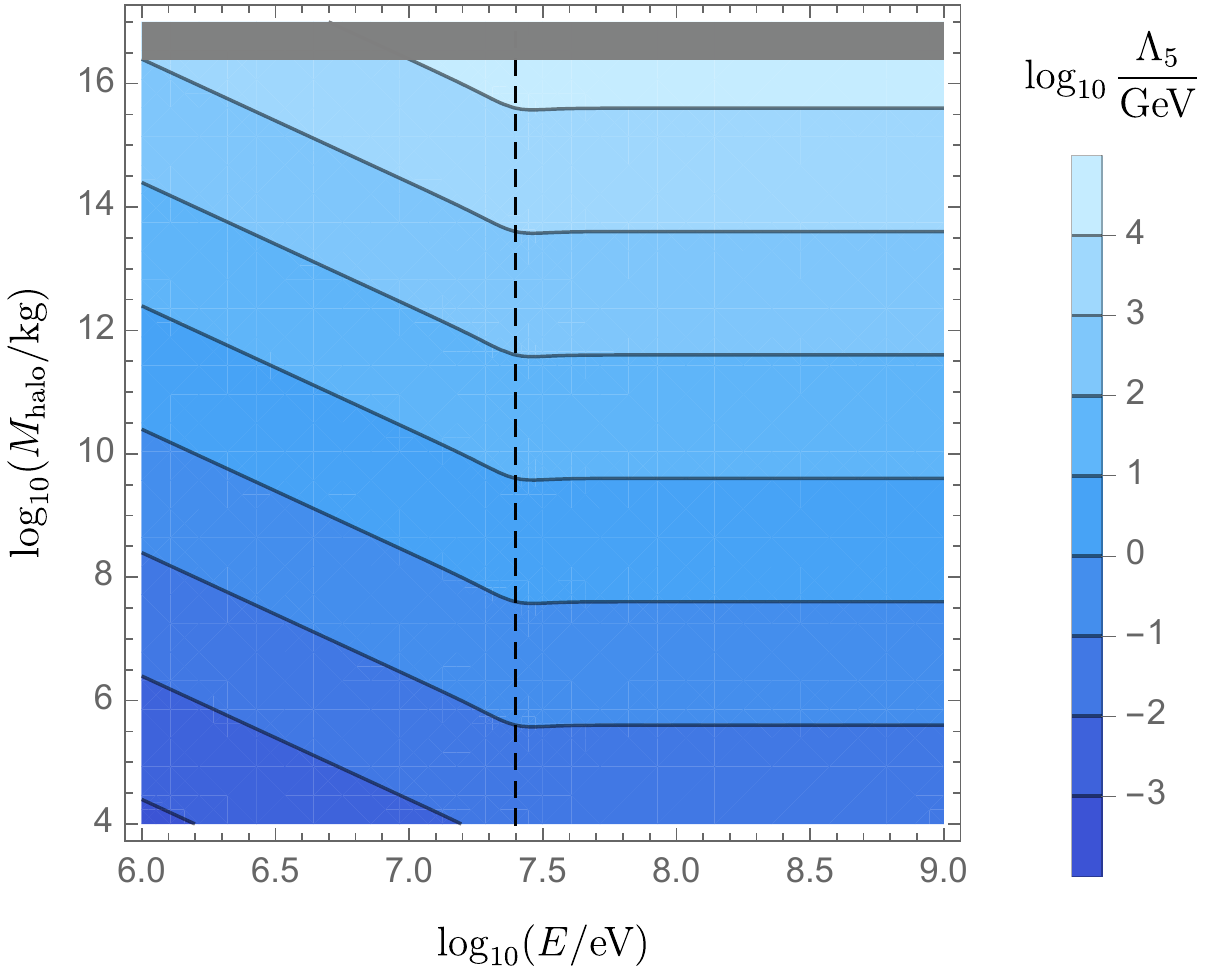}}
		\end{minipage}
	}
\caption{\textit{Left:} Contours showing the value of the scalar-neutrino coupling $y$ in (\ref{L_int2}), for a given neutrino energy $E$ and halo mass $M_{\text{halo}}$, at which the relative deviation from the vacuum neutrino oscillation probability is $0.1$.
\textit{Right:} A similar contour plot for the suppression scale $\Lambda_5$ of the scalar-neutrino interaction (\ref{L_int1}).
We assume $m=10^{-10}$ eV (big halo), $\Delta m_0^2=2.5\times 10^{-3}$ eV$^2$, and $\sin^22\theta_0=0.99$.
The gray shaded region depicts the experimentally excluded values of  $M_{\text{halo}}$ and the vertical dashed line depicts the value of $E$ at which $\eta=1$ (see \cref{eta}).
}
\label{fig:sens}
\end{figure*}

\begin{figure}[b]
\center{
\includegraphics[width=0.5\linewidth]{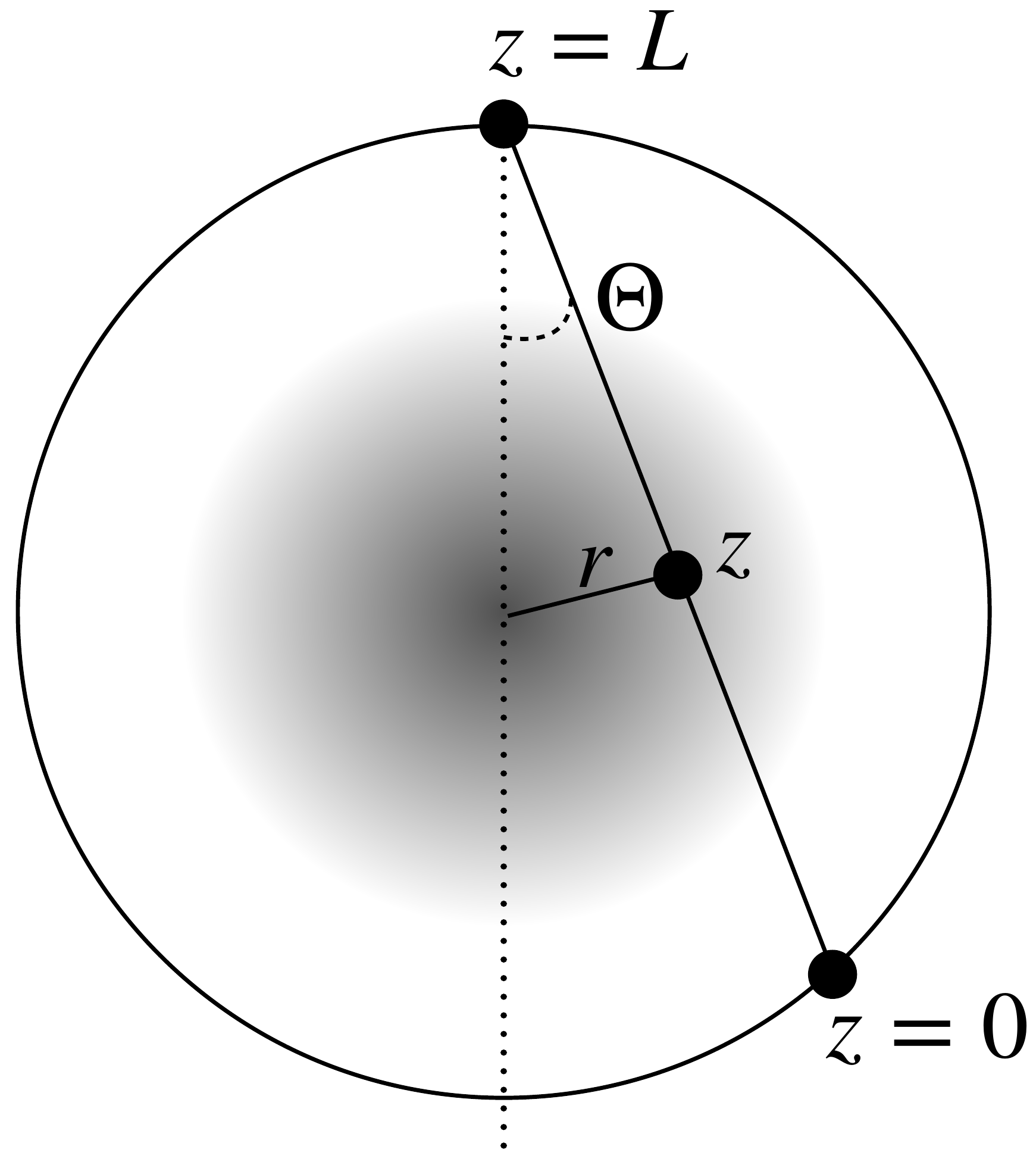}
\caption{The path of a neutrino propagating through a halo core located inside the Earth.}
\label{fig:geom}}
\end{figure}

We next turn to the dimension-four interaction (\ref{L_int2}).
The important difference is the presence of the quadratic $\vf$ term in the Hamiltonian (\ref{DeltaH1}).
This term dominates the linear $\vf$ term when $\epsilon\gtrsim\eta^{-1}(\sum m_\nu)/\sqrt{\Delta m_0^2}$.
On the other hand, repeating the computation of the correction $\Delta\nu_i(z)$ to the mass eigenstates, we obtain that the quadratic term results in the following correction
\begin{equation}
    |\Delta\nu_i(L_{0}^{\rm osc})|\sim \epsilon^2\eta^2 \frac{\Delta m_0^2}{\left(\sum m_\nu\right)^2}  |\tilde{h}_{ij}|\;,
\end{equation}
where $\Tilde{h}=U_0^\dagger h^\dagger h U_0$.
We see that, barring the order-one ratio $\sqrt{\Delta m_0^2}/\sum m_\nu\gtrsim 0.5$, the parameter governing the nonadiabatic oscillations in the presence of the interaction (\ref{L_int2}) is $\epsilon\eta$, which is similar to the adiabatic regime.
To confirm this, we solve numerically \cref{EvolEq} with the Hamiltonian (\ref{H0}), (\ref{DeltaH2}), and compute the survival probability averaged over the halo phase.
The result is shown in Fig.~\ref{fig:osc_nonad4}, where we take again $E=1$ GeV and several values of $\epsilon\eta$. 
Note that the quadratic $\vf$ term in the Hamiltonian (\ref{DeltaH2}) does not induce secondary oscillations, unlike the linear term, and the latter are suppressed.
In the limit $\epsilon\eta\gg 1$, the averaged survival probability tends to $1/2$.

In summary, for neutrinos with $\eta\ll 1$, the small parameter controlling the deviation from the vacuum oscillations is $\epsilon\eta$, and the effect of the halo background is similar to that of a homogeneous matter potential.
For neutrinos with $\eta\gg 1$, the small parameter is $\epsilon$ in the case of the derivative interaction (\ref{L_int1}), and $\epsilon\eta$ in the case of the marginal interaction (\ref{L_int2}).
Using the definitions (\ref{epsilon}), (\ref{eta}), these results can be rephrased in terms of the neutrino energy, the halo mass, and the scalar-neutrino coupling parameter, $y$ or $\Lambda_5$.
This is done in Fig.~\ref{fig:sens}, which shows the values of $y$ (in the dimension-four interaction (\ref{L_int2})) or $\Lambda_5$ (in the dimension-five interaction (\ref{L_int1})) necessary for a big halo composed of particles with $m=10^{-10}$ eV to induce a $10\%$ deviation in the oscillation probability, at a given neutrino energy and a given halo mass.
Depending on the choice of the scalar-neutrino interaction term, the sensitivity to the halo is either energy independent (for $y$), or reaches its maximum at $E\gtrsim 25$ MeV (for $\Lambda_5$).
Thus, neutrinos interacting with sufficiently heavy scalar halos constrain dimensionless couplings to be $y\lesssim 10^{-16}$ and dimension-five scales $\Lambda_5\lesssim 10^5$~GeV.

\subsection{Probing the small (interior) halo}
\label{ssec:sub}

For a scalar mass $m\gtrsim 10^{-9}$ eV, the halo core is located inside the Earth.
Such a halo can be probed by neutrinos traversing the Earth, with their source and detector located on opposite sides of the planet (see Fig.~\ref{fig:geom} for an illustration).
This setup is typical for atmospheric neutrinos, and we will consider oscillation parameters relevant for a GeV-scale neutrino: $\Delta m_0^2\approx 2.5\times 10^{-3}$ eV, $\sin^22\theta_0\approx 0.087$ \cite{ParticleDataGroup:2022pth}. 
Note that, depending on the neutrino mass ordering and energy, the contribution to the Hamiltonian (\ref{H0}) generated by the Earth's matter can significantly affect atmospheric neutrino oscillations and enhance them resonantly (see \cite{Akhmedov:2006hb} and references therein).
For simplicity, we do not consider this matter effect here. 
This is justified since, unlike the Earth's matter, the resonant oscillations due to the halo can occur for neutrinos in a broad range of energies, as seen in Section~\ref{sssec:nonad}.

\begin{figure}[t]
\center{
\includegraphics[width=0.9\linewidth]{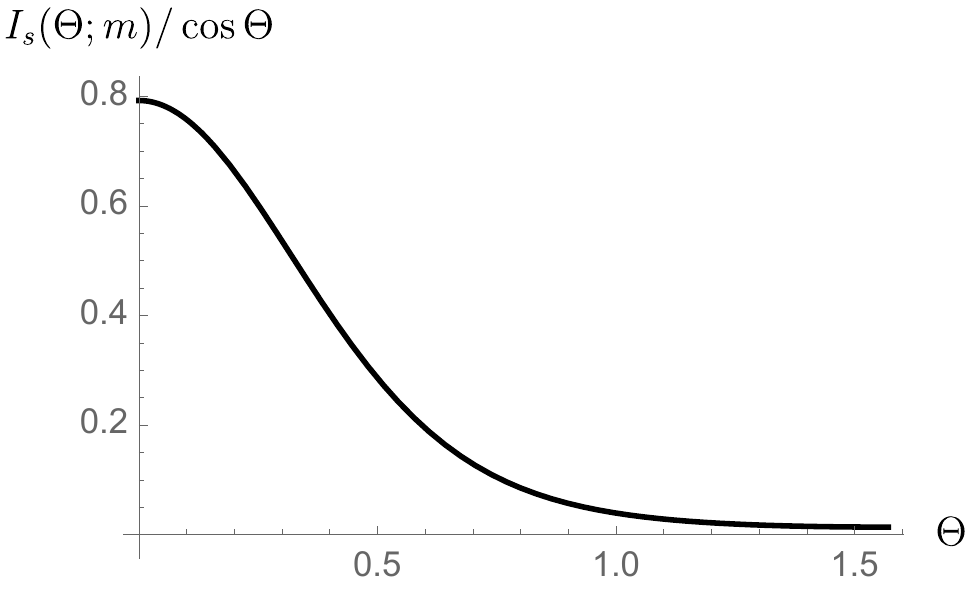}
\caption{The angle-dependent part of the correction in \eqref{DeltaMEff2} to the neutrino mass-squared difference as a function of the nadir angle of the incoming neutrino, for $m=5\times 10^{-9}$ eV.}
\label{fig:I(psi)}}
\end{figure}

First, we repeat the analysis of the perturbative corrections in the adiabatic regime. Using \cref{l_m_2} it is convenient to rewrite the parameters (\ref{epsilon}), (\ref{eta}) as
\begin{equation}\label{epsilon1}
\epsilon \sim \left(\frac{\beta}{10^{-23}}\right)\left(\frac{10^{-9}\:\text{eV}}{m}\right)^{5/4}\left( \frac{M_{\text{halo}}}{10^{15}\:\text{kg}} \right)^{1/2} \;,
\end{equation}
\begin{equation}\label{eta1}
\eta=400 \left(\frac{2.5\times 10^{-3}\:\text{eV}^2}{\Delta m_0^2}\right)\left(\frac{m}{10^{-9}\:\text{eV}}\right)\left(\frac{E}{1\:\text{GeV}}\right) \;.
\end{equation}
Here the parameter $\epsilon$ contains the amplitude of the field $f_0$ in the center of the halo.
Next, for atmospheric neutrinos one clearly obtains $\eta\gg 1$.
Additionally, $mL\gg 1$ for neutrinos traversing the Earth.
This allows us to compute the effective mass-squared difference $\Delta m_{\text{eff}}^2$ in \cref{P_surv} independently of the rest of the probability.
Furthermore, it is convenient to express $\Delta m_{\text{eff}}^2$ as a function of the nadir angle $\Theta$ of the incoming neutrino.
From \cref{DeltaMEff,ParamExp2,corr2,Gamma} we obtain
\begin{equation}\label{DeltaMEff2}
\Delta m_{\text{eff}}^2=\Delta m_0^2\left( 1+2\epsilon^2\eta^2 \A_{m2} \frac{I_s(\Theta;m)}{\cos\Theta} \right) \;,
\end{equation}
where we define
\begin{equation}
    I_s(\Theta;m)\equiv\limitint_0^{2s}\diff x\, \widehat{f}^2(\sqrt{x^2+1-2xs}) \;,
\end{equation}
$s=\cos\Theta$, $x=z/R_{\oplus}$, and $\widehat{f}$ is the normalised halo profile, $\widehat{f}(0)=1$.
The function $I_s(\Theta;m)/\cos\Theta$ is plotted in Fig.~\ref{fig:I(psi)} for the scalar mass $m=5\times 10^{-9}$ eV at which the halo size is comparable to that of the Earth, $\ell\approx 0.5 R_{\oplus}$.

\begin{figure}[b]
\center{
\includegraphics[width=0.95\linewidth]{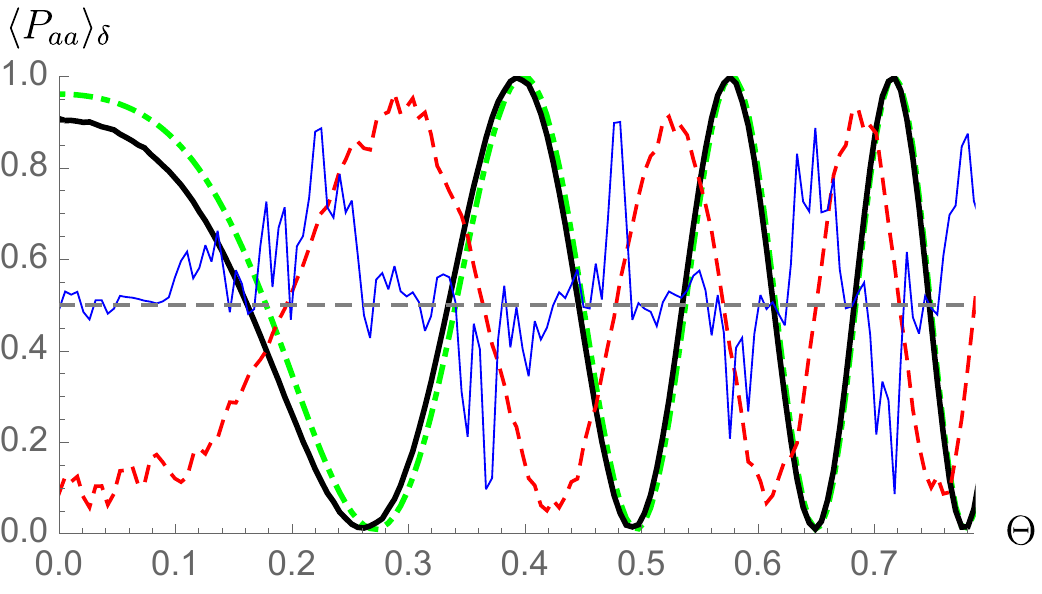}
\caption{The survival probability of the neutrino traversing the halo inside the Earth, averaged over the halo phase, for several values of $\epsilon$ \eqref{epsilon1}.
We assume the derivative scalar coupling (\ref{L_int1}), $m=3\times 10^{-9}$ eV, $E=1$ GeV, and the vacuum oscillation parameters $\Delta m_0^2\approx 2.5\times 10^{-3}$ eV, $\sin^22\theta_0\approx 0.99$.
When $\epsilon=0.1$ (black line), the deviation from the vacuum oscillation (green, dot-dashed line) is only sizeable at small nadir angles $\Theta$ corresponding to the neutrino traversing the halo core (see fig.~\ref{fig:geom}). 
When $\epsilon=0.5$ (red, dashed line), the deviation is visible at all $\Theta$. 
For even higher values, $\epsilon=1.5$ (thin, blue line), the probability tends to $1/2$ (gray, dashed line).
All probabilities are plotted with the step $\Delta\Theta=0.01$.
}
\label{fig:SmallHaloTheta}}
\end{figure}

\begin{figure}[t]
 \center{
 \includegraphics[width=0.95\linewidth]{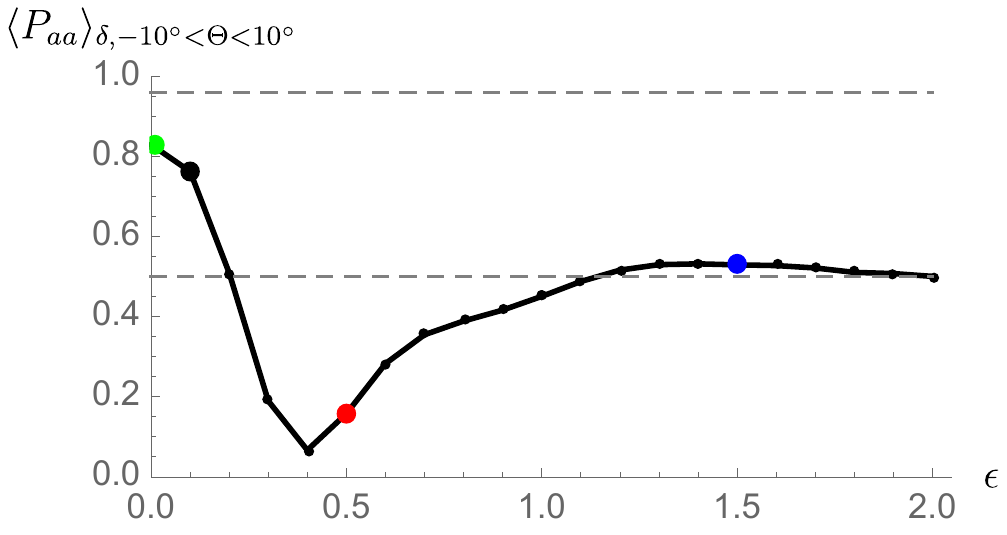}
 \caption{The survival probability of the neutrino traversing the halo inside the Earth, averaged over the halo phase and the incoming angle $\Theta$ in the range $-10^\circ<\Theta<10^\circ$, as a function of $\epsilon$ (\cref{epsilon1}).
 We assume the derivative scalar coupling (\ref{L_int1}), $m=3\times 10^{-9}$~eV, $E=1$ GeV, and the vacuum oscillation parameters $\Delta m_0^2\approx 2.5\times 10^{-3}$ eV, $\sin^22\theta_0\approx 0.99$.
 As $\epsilon$ grows, the probability interpolates between its vacuum value and $1/2$ (gray dashed lines).
 The big colored dots correspond to the curves shown in Fig.~\ref{fig:SmallHaloTheta}.
 }
 \label{fig:SmallHaloEps}}
\end{figure}

\begin{figure*}[t]
    \center{
        \begin{minipage}[h]{0.45\linewidth}
            \center{\includegraphics[width=0.99\linewidth]{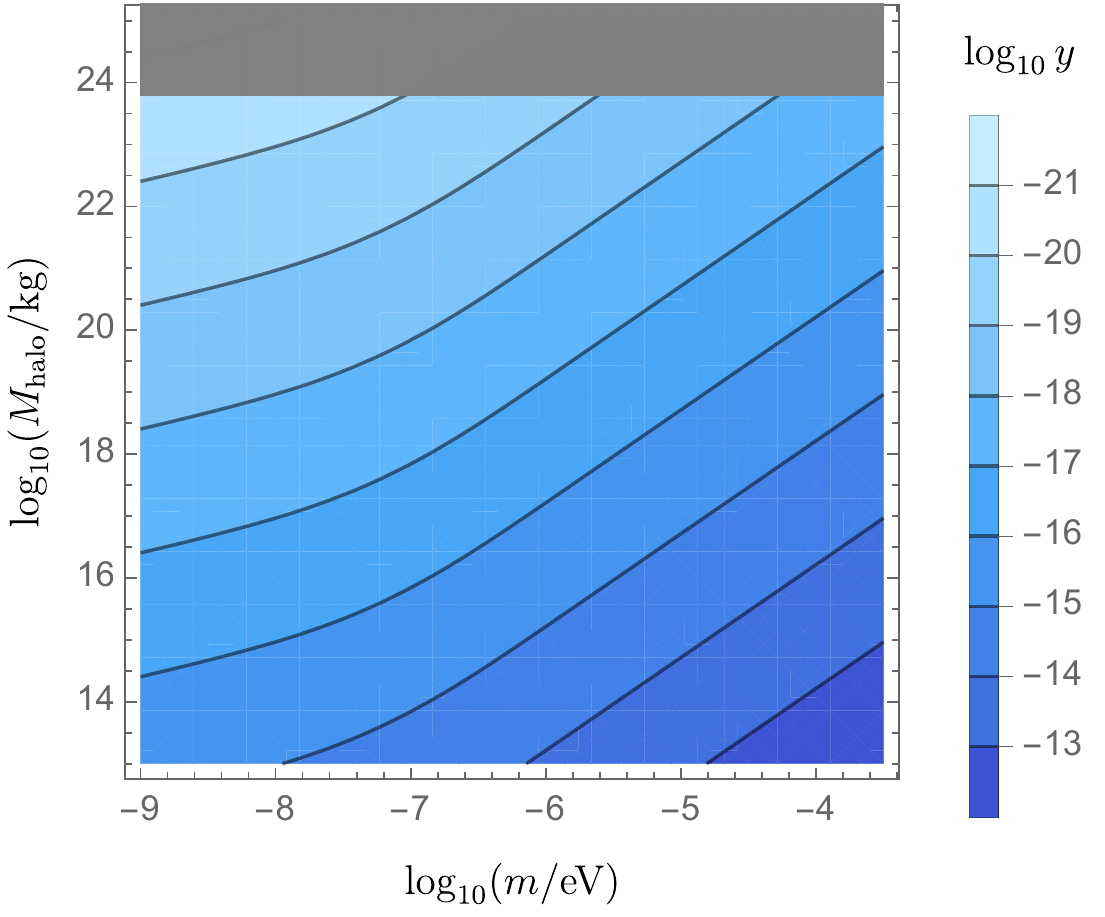}}
        \end{minipage}
		\hfill
		\begin{minipage}[h]{0.45\linewidth}
			\center{\includegraphics[width=0.99\linewidth]{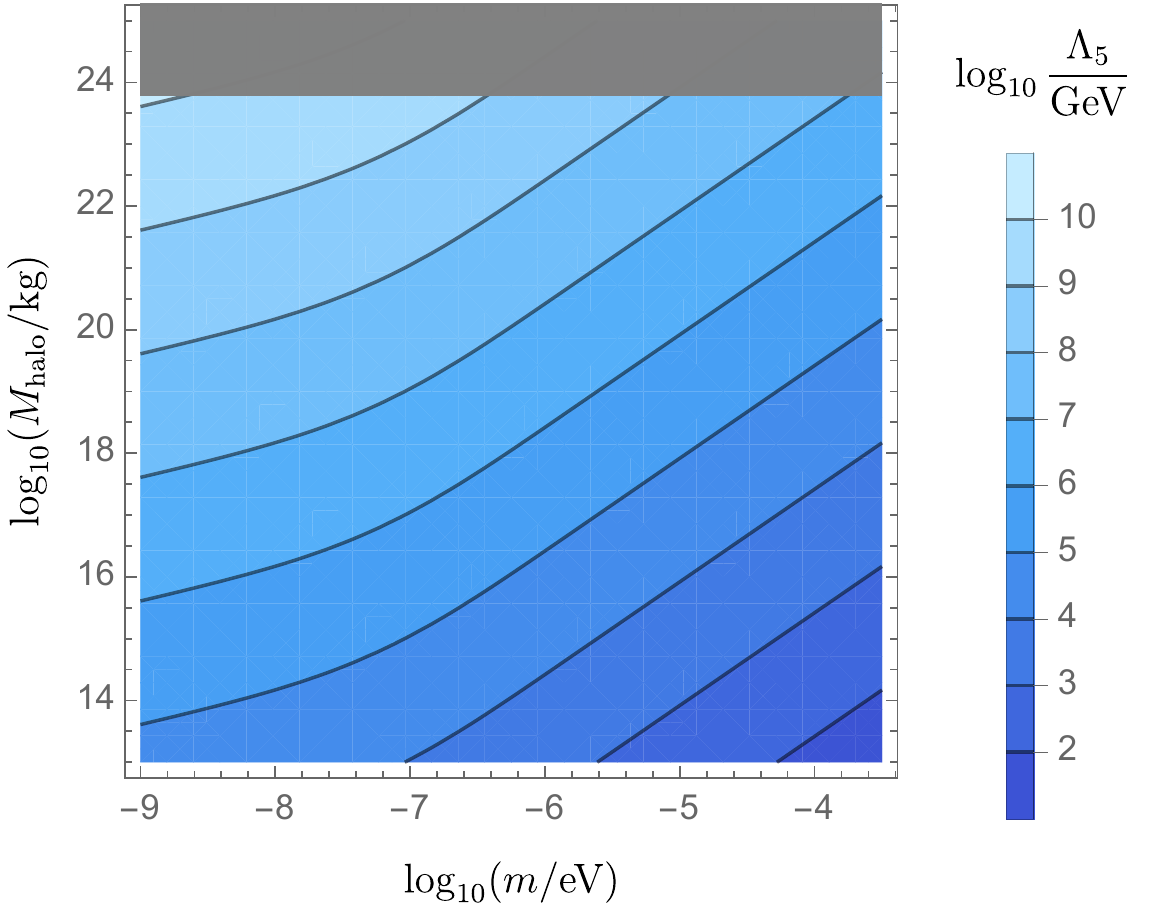}}
		\end{minipage}
	}
\caption{\textit{Left:} Contours showing the values of the scalar-neutrino coupling $y$ in (\ref{L_int2}) for a given scalar field mass $m\gtrsim 10^{-9}$ eV (small halo) and halo mass $M_{\text{halo}}$, at which the relative deviation from the vacuum neutrino oscillation probability is $0.1$.
\textit{Right:} A similar contour plot for the suppression scale $\Lambda_5$ of the scalar-neutrino interaction (\ref{L_int1}).
We assume the angular resolution of the neutrino detector $\Theta_{\text{res}}=20^\circ$, the neutrino energy $E=1$ GeV, $\Delta m_0^2=2.5\times 10^{-3}$ eV$^2$, and $\sin^22\theta_0=0.99$.
The gray shaded region depicts the values of $M_{\text{halo}}>0.1 M_{\oplus}$.
}
\label{fig:sensSmall}
\end{figure*}

In a realistic setup, due to the limited angular and energy resolution of a neutrino detector, one is sensitive to the oscillation probability which is averaged over the position of the neutrino source and the neutrino energy band $\Delta E\gg m$.
The averaged probability can be written as
\begin{equation}
\langle P_{aa}\rangle_{\delta,L,\Delta E}=1-\frac{1}{2}\sin^22\theta_{\text{eff}} \;.
\end{equation}
Using \cref{P_surv,ParamExp1,corr1,Gamma}, we obtain the effective mixing angle,
\begin{equation}\label{ThetaEff}
\theta_{\text{eff}}=\theta_0+2\epsilon_{\oplus}^2\eta^2(\A_{\theta 2}+\A_{\theta 1}^2\cot 2\theta_0) \;,
\end{equation}
where $\epsilon_{\oplus}=\epsilon f_{\oplus}/f_0$, and $f_{\oplus}$ is the amplitude of the halo at the Earth's surface.
Clearly, the correction to the mixing angle is additionally suppressed by a factor $f^2_{\oplus}/f^2_0$ compared with the mass-squared difference \eqref{DeltaMEff2}.

As discussed in Section~\ref{sssec:nonad}, the halo time variation severely limits the applicability of the adiabatic approximation for neutrinos with $\eta\gg 1$, and the corrections (\ref{DeltaMEff2}), (\ref{ThetaEff}) are only valid for $\epsilon\ll\eta^{-2}$, which, by \cref{eta1}, limits the deviation from vacuum oscillation of neutrino with $E\sim 1$ GeV to be $\lesssim 1\%$. 
When $\epsilon\gtrsim\eta^{-2}$, one needs to solve \cref{EvolEq} numerically.
From the results of the previous section one can nevertheless draw a qualitative picture of what happens at larger values of $\epsilon$.
Namely, the correction to the oscillation probability due to the halo is expected to be small for all incoming neutrinos until the amplitude of the field in the halo center is such that $\epsilon\sim\eta^{-1}$ (for the interaction (\ref{L_int2})) or $\epsilon\sim 1$ (for the interaction (\ref{L_int1})).
Furthermore, if the halo mass increases, the largest value of the nadir angle $\Theta$ at which the oscillation probability is significantly affected by the halo also increases.

As an illustration, Fig.~\ref{fig:SmallHaloTheta} shows an example of the numerical calculation of the survival probability, averaged according to \cref{P_surv_av}, as a function of $\Theta$, where, for concreteness, we choose the interaction (\ref{L_int1}).
We also take the scalar field mass $m=3\times 10^{-9}$ eV, corresponding to the halo size close to the size of the Earth, $\ell\approx 0.65 R_{\oplus}$.
We see indeed that the magnitude of the deviation from the vacuum probability is controlled by the parameter $\epsilon$.
For $\epsilon\ll 1$, the effect may only be visible at small $\Theta$, when the neutrino passes through the core of the halo, owing to the fact that $\ell/L_0^{\rm osc}\sim 10$ for $\ell\sim R_{\oplus}$.
In the opposite regime, $\epsilon\gg 1$, the probability tends to $1/2$ irrespective of the angle.

Fig.~\ref{fig:SmallHaloEps} shows the survival probability averaged over the incoming angle $\Theta$ in the range $-10^\circ<\Theta<10^\circ$, for the same parameters as in Fig.~\ref{fig:SmallHaloTheta}. 
We see that a $10$\% deviation from the vacuum probability is achieved at $\epsilon\approx 0.1$; at $\epsilon\gtrsim 1$ the probability becomes close to $1/2$.

What happens at much larger values of $m$ corresponding to much smaller halos?
Assume that the neutrino detector has a certain angular resolution $\Theta_{\text{res}}$.
By smearing \eqref{epsilon1} over $\Theta_{\text{res}}$ one can define an effective expansion parameter:
\begin{equation}\label{epsilonEff}
    \epsilon_{\text{eff}} = \Theta_{\text{res}}^{-1}\limitint_0^{\Theta_{\text{res}}}\diff\Theta\: \epsilon(\Theta) \;,
\end{equation}
where
\begin{equation}
    \epsilon(\Theta) = \frac{\beta}{m} f(R_{\oplus}\sin\Theta)\,,
\end{equation}
corresponds to the maximal amplitude of the halo probed by the neutrino with the angle $\Theta$. 
At a given $m$ and $E$, one can infer the halo mass corresponding to, e.g., $\epsilon_{\text{eff}}\eta= 0.1$ (for the interaction (\ref{L_int2})) or $\epsilon_{\text{eff}}= 0.1$ (for the interaction (\ref{L_int1})).
The result is shown in Fig.~\ref{fig:sensSmall}, where, for concreteness, we take $\Theta_{\rm res}=20^\circ$.
We see that the absence of the constraint on $M_{\text{halo}}$ from the lunar laser ranging allows us to probe much lower values of $y$ (or higher values of $\Lambda_5$). In particular, for $m\sim 10^{-9}$ eV, couplings as small as $10^{-21}$ and scales as large as $10^{10}$ GeV can be probed. 
However, the sensitivity diminishes as the scalar field mass increases since this corresponds to decreasing the halo size, which then contributes less to the integral in \cref{epsilonEff}.
Also, changing $\Theta_{\rm res}$ leads to a proportional change in the sensitivity.

\section{Local vector halo}
\label{sec:vector}

\subsection{Nonrelativistic vector soliton}

In this section we repeat the analysis in the previous sections for the case when the halo is made of massive vector particles, such as a dark photon.
Coupling the $U(1)$ vector field to the neutrino current leads to new effects in the neutrino oscillations due to the polarisation \cite{Brdar:2017kbt,Capozzi:2018bps,Brzeminski:2022rkf}.
We again assume the Earth hosts the halo, but this time arising from a $U(1)$ massive vector field $A_\mu$.
To obtain a soliton solution, we consider radially polarised, spherically symmetric configurations described by the ansatz (see also~\cite{Loginov:2015rya})
\begin{equation}\label{ansatz_vec}
    A_t(r,t)=c u(r)\cos\omega t \;, ~~~ A_r(r,t)=v(r)\sin\omega t \;,
\end{equation}
and $A_\theta=A_\phi=0$.
In the gravitational background (\ref{ds}), the equations of motion for the components $u$, $v$ are
\begin{subequations}\label{eom4}
\begin{align}
    & \omega v(r)-cu'(r)=\frac{m^2c^4}{\omega}N(r) v(r) \;, \label{eom4a} \\ 
    & \frac{1}{r^2}\frac{\diff }{\diff r}(r^2(cu'(r)-\omega v(r)))=m^2 c^3 \frac{u(r)}{N(r)} \;, \label{eom4b}
\end{align}
\end{subequations}
where $m$ is the mass of the vector boson.
These equations are analogous to those appearing in the studies of self-gravitating, relativistic, (complex) vector field configurations -- Proca stars \cite{Brito:2015pxa}. 
The important difference is, however, that in our case the function $N(r)$ is fixed by the background metric.

Since the equation of motion does not contain the second time derivative of $A_t$, the dynamical degree of freedom is associated with the function $v(r)$.
Nevertheless, it is convenient to write \cref{eom4} as a differential equation on $u(r)$.
Taking the nonrelativistic limit $\omega=mc^2+E$, with $|E|\ll  mc^2$, and using the units (\ref{units}), we obtain, to leading order in $c$,
\begin{subequations}\label{eom5}
\begin{align}
        -\frac{1}{x^2}\frac{\diff }{\diff x}\left(x^2\frac{\diff u}{\diff x}\right)&+\frac{\M\tilde{\Phi}'}{\M\tilde{\Phi}-\E}\frac{\diff u}{\diff x}+2(\M\Tilde{\Phi}-\mathcal{E})u =0 \;, \label{eom5a} \\
        & v=\frac{mR_{\oplus}}{2(\E-\M\tilde{\Phi})} \frac{\diff u}{\diff x}\;, \label{eom5b}
\end{align}
\end{subequations}
where $\tilde{\Phi}$ is given in \cref{PhiTilde}.
Unlike the scalar field \cref{eom2}, there is now a ``friction'' term in \cref{eom5a}.
This term becomes negligible in the limits $x\to 0$ and $x\to\infty$, in which one recovers the scalar wavefunction.

\begin{figure}[t]
\center{
\includegraphics[width=0.85\linewidth]{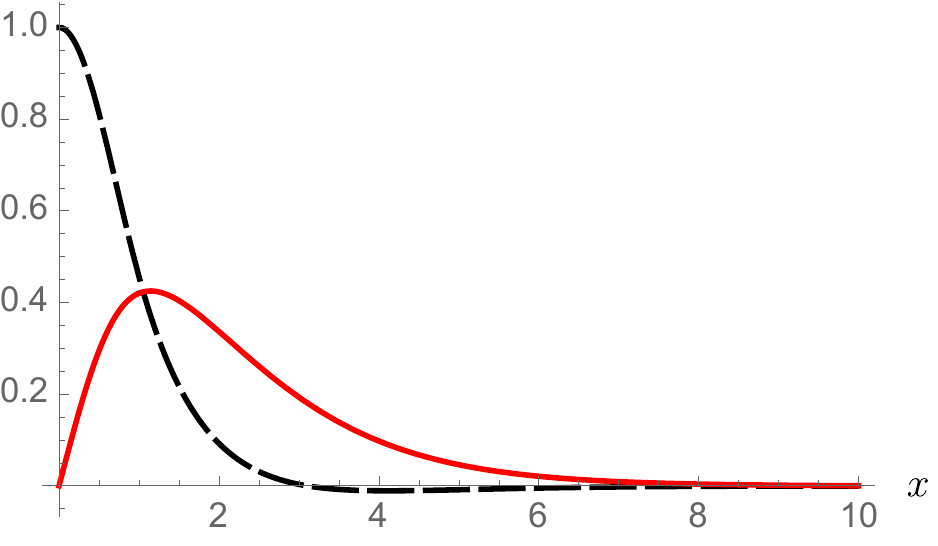}
\caption{A ground-state solution of the vector field equation of motion \eqref{eom5}, assuming $\M=2$. 
The black dashed line depicts the normalised function $u_0(x)$, while the red line depicts $v_0(x)/(mR_{\oplus})$.}
\label{fig:f,g}}
\end{figure}

The equation of motion \eqref{eom5a} is solved numerically for the ground state $u_0(x)$ where the usual boundary conditions of regularity at the origin and vanishing at infinity are imposed. A particular ground-state solution $u_0(x)$, $v_0(x)$ is shown in Fig.~\ref{fig:f,g}.
Note that the vector halo profile is not monotonic in $x$; in particular, $u_0'$ vanishes together with $\M\Tilde{\Phi}-\E_0$ at a finite value of $x$.
It is reasonable to use this value as the size of the vector halo $x_\ell$.
The vector halo size exhibits the same asymptotic behaviour as the scalar halo \eqref{x_st} (up to order-one factors), and therefore the estimates (\ref{l_m}) remain valid in the vector case.

\subsection{Vector-neutrino coupling}

We turn to the vector-neutrino coupling and consider the following interaction
\begin{equation}\label{L_int3}
\L_{V,\rm int}=-y_V \kappa_{ab} A_\mu \bar{\psi}_{La}\gamma^\mu\psi_{Lb} \;,
\end{equation}
where $y_V$ is a small dimensionless coupling and the Hermitian coupling matrix, $\kappa_{ab}$ has order one matrix elements. 
We are agnostic about a particular model generating this interaction; as an example, $A_\mu$ can be identified with the $U(1)_{L_e-L_\mu}$ and $U(1)_{L_\mu-L_\tau}$ gauge bosons that give rise to a flavour non-universal vector-neutrino coupling \cite{Dror:2019uea,Fabbrichesi:2020wbt,Brzeminski:2022rkf}.

The main difference between the neutrino interaction (\ref{L_int3}) with the radially-polarised vector halo and the derivative interaction (\ref{L_int1}) with the scalar halo is that in the former case the spatial component of the neutrino current dominates 
for the range of vector field masses we are interested in.
Indeed, using \cref{l_m,eom5b}, we obtain $|v_0|\gg|u_0|$ provided $10^{-12}\,{\rm eV} \lesssim m \lesssim 10^{-3}$ eV.

The interaction (\ref{L_int3}) modifies the neutrino dispersion relation by shifting the neutrino momentum.
Assuming $E\gg y_V|v_0|$ along the neutrino trajectory, one obtains the following contribution to the Hamiltonian in the flavour basis,
\begin{equation}\label{H2}
\Delta H_V = | E\vec{n}\mathbf{1}+y_V\vec{A}\kappa|-E\mathbf{1} \;,
\end{equation}
where $\vec{n}$ is a unit vector in the direction of neutrino propagation and $\mathbf{1}$ is the $2\times2$ identity matrix.
If one further assumes that $y_V|v_0|\ll\Delta m_0^2/E$, then \cref{H2} simplifies to
\begin{equation}\label{H3}
    \Delta H_V=y_V \vec{n}\cdot\vec{A}\: \kappa \;.
\end{equation}
Comparing with the derivative coupling to the scalar halo \eqref{DeltaH1}, we see that the perturbative analysis of Section~\ref{sec:neutrino} readily applies to the vector halo. 
The perturbative parameter is now $\beta=y_V$, and the function $\Gamma(z)$ is defined as 
\begin{equation}\label{Gamma2}
 \Gamma(z)=2E\vec{n}\cdot\vec{A}=2E\left(z-\frac{L}{2}\right)\frac{v_0(r(z))}{r(z)}\cos (mz+\delta) \;.
\end{equation}
We work in the coordinate system shown in Fig.~\ref{fig:geom} where $r^2(z)=z^2+R_{\oplus}^2-2zR_{\oplus}\cos\Theta$.
Finally, the parameters $\A_{\theta,m}$ are given in \cref{As11,As12,As21,As22,As23,As24} of Appendix~\ref{sec:AppA}, with $g_{ab}$ replaced by $\kappa_{ab}$.

\begin{figure}[b]
\center{
\includegraphics[width=0.9\linewidth]{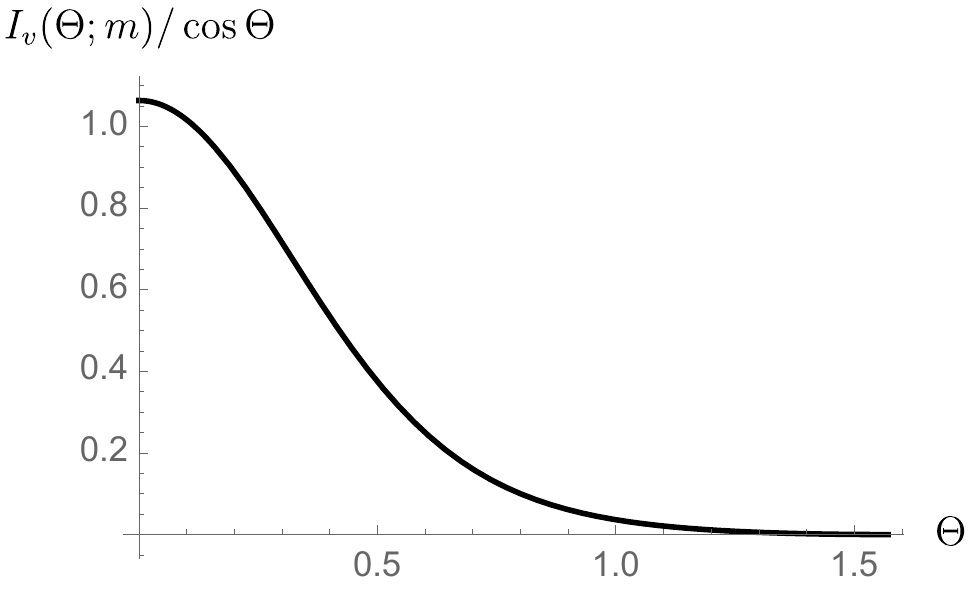}
\caption{The angle-dependent part of the vector halo correction in \eqref{DeltaMEff3} to the neutrino mass-squared difference 
as a function of the nadir angle of the incoming neutrino, assuming $m=5\times 10^{-9}$ eV.}
\label{fig:I(psi)_v}}
\end{figure}

\begin{figure}[t]
\center{
\includegraphics[width=0.85\linewidth]{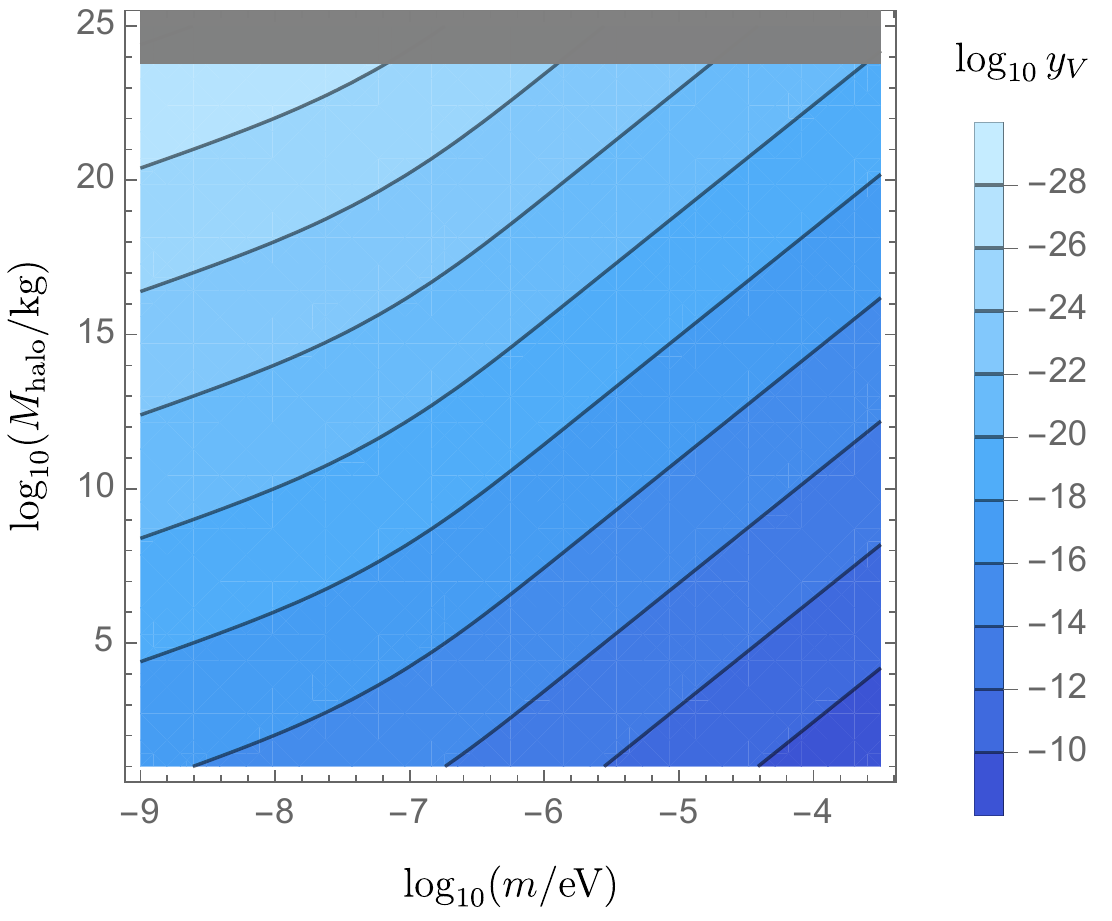}
\caption{Contours showing the values of the vector-neutrino coupling $y_V$ in (\ref{L_int3}) for a given vector field mass $m\gtrsim 10^{-9}$ eV (small halo) and halo mass $M_{\rm halo}$, at which the relative deviation from the vacuum neutrino oscillation probability is $0.1$.
We assume the angular resolution of the neutrino detector is $\Theta_{\text{res}}=20^\circ$, the neutrino energy $E=1$ GeV, $\Delta m_0^2=2.5\times 10^{-3}$ eV$^2$, and $\sin^22\theta_0=0.99$.
The gray shaded region depicts the values of $M_{\text{halo}}>0.1 M_{\oplus}$.}
\label{fig:sensV}}
\end{figure}

For the big halo, due to its radial polarisation, terrestrial experiments involving reactor or accelerator neutrinos with baselines $L\ll R_{\oplus}$ are less advantageous than in the scalar case, since the neutrino propagates almost orthogonally to the vector $\vec{A}$.
However for the small halo, the effects can be much larger and it is straightforward to derive the correction to the mass-squared difference (cf. \cref{DeltaMEff2}),
\begin{equation}\label{DeltaMEff3}
\Delta m_{\text{eff}}^2=\Delta m_0^2\left( 1+\frac{1}{2}\epsilon^2\eta^2 \A_{m2}\frac{I_v(\Theta;m)}{\cos\Theta} \right) \;,
\end{equation}
where $f_0$ is replaced by $v_0(0)$ in the definition of $\epsilon$ (\ref{epsilon}),
\begin{equation}
I_v(\Theta;m)\equiv\limitint_0^{2s}\diff x\:\frac{(x-s)^2}{x^2+1-2xs}\, \widehat{v}_0^2(\sqrt{x^2+1-2xs})\;,
\end{equation}
and $\widehat{v}_0$ is normalised so that $\widehat{v}_0=1$ at the maximum.
The function $I_v(\Theta;m)/\cos\Theta$ is shown in Fig.~\ref{fig:I(psi)_v} for the vector mass $m=5\times 10^{-9}$ eV. The function is monotonic even though the radial component of the halo profile $v_0$ is not.
The effective mixing angle is given by \cref{ThetaEff} where now $\epsilon_{\oplus}=\epsilon v_0(R_{\oplus})/v_0(0)$.
Note that the condition $y_V|v_0|\ll\Delta m_0^2/E$ made to simplify the Hamiltonian (\ref{H2}) is equivalent to the condition $\epsilon\eta\ll 1$ corresponding to the validity of perturbation theory.

When the adiabatic condition (\ref{CondAd}) is violated, the equation of motion \eqref{EvolEq} with the Hamiltonian (\ref{H0}), (\ref{H2}) must be solved numerically.
Let us focus on the most interesting case of a small halo for which $m\gtrsim 10^{-9}$ eV.
We introduce again the parameter $\epsilon_{\rm eff}$ via \cref{epsilonEff}, where now $\epsilon(\Theta)= y_V v_0(R_{\oplus}\sin\Theta)/m$, and compute the halo mass corresponding to $\epsilon_{\rm eff}=0.1$, that is, to $\sim 10$\% relative difference between the measured neutrino survival probabilities in the presence and absence of the halo.
The result is shown in Fig.~\ref{fig:sensV}.
It closely resembles the plot on the right panel of Fig.~\ref{fig:sensSmall} upon changing the variable $\Lambda_5\mapsto m/(2y_V)$.
This is because of the similarity between the neutrino coupling (\ref{L_int3}) to the radially-polarised vector halo and the derivative coupling (\ref{L_int2}) to the scalar halo: as soon as $E\gg m$, both types of interactions lead to the linear in $\vf$ (or $A_\mu$) and energy-independent correction to the neutrino Hamiltonian.

\section{Conclusion}
\label{sec:concl}

The background potential of massive, astrophysical objects can bound dark matter particles (scalars or vector bosons) to form a halo surrounding the object. 
Assuming that the possible dark matter interactions, either self or with ordinary matter, play no role in sustaining the halo, one can analytically solve for the nonrelativistic, solitonic halo configuration in a spherically-symmetric gravitational potential due to the host body. 
For the Earth as the host body, the halo extends beyond the Earth's surface and remains homogeneous for $10^{-10}$~eV~$\lesssim m\lesssim 10^{-9}$~eV, while for $m\gg 10^{-9}$~eV the halo forms within the Earth's interior.
Furthermore, the density of dark matter in the halo can be much larger than the average relic density, thereby increasing the possibility to detect it.

An interesting way to detect such a local dark matter halo is to assume that it interacts with the neutrino. There are several dark matter-neutrino interactions that modify the neutrino dispersion relation and assuming two-flavour oscillations and neglecting the MSW effect, we computed the distortions of the vacuum oscillations caused by the halo. The corresponding survival probability can be calculated analytically in perturbation theory and within the adiabatic approximation. 
Beyond the domain of validity of the adiabatic approximation, the survival probability was computed numerically.
Nonadiabatic effects manifest themselves as multiple resonances during the neutrino propagation, caused by the halo time variation.
Despite the resonances, the deviation from the vacuum oscillation probability can still be small if the halo remains sufficiently light.

For dark matter masses $m\gtrsim 10^{-10}$ eV, one cannot rely on the periodic modulation of neutrino parameters that follows from the time variation of the coherent dark matter background (see, e.g., \cite{Krnjaic:2017zlz}).
Instead, the correction to the oscillation probability is due to the enhanced dark matter density in the halo.
We showed that for neutrino energies ($E/25$ MeV)$\gg$($10^{-9}$~eV$/m$)$\times$($\Delta m_0^2/2.5\times 10^{-3}$ eV$^2$), corresponding to $\eta\gg 1$, the visible correction is driven by the nonadiabatic effects.
Depending on the halo mass, these effects can alter observably the average oscillation probability for a broad range of energies, including those typical for accelerator and atmospheric neutrinos.

Since there is no strong bound on the total mass of dark matter, which can possibly be accumulated inside the Earth, assuming a sufficiently heavy local dark matter halo puts much stronger constraints on the dark matter-neutrino interactions than other terrestrial, solar, astrophysical, or cosmological considerations (see, e.g., \cite{Krnjaic:2017zlz,Brdar:2017kbt,Dev:2020kgz,Dev:2022bae}).
In the big halo case, corresponding to $m\ll 10^{-9}$~eV, we are able to probe the scalar-neutrino Yukawa-like coupling (\ref{L_int2}) down to $10^{-15}$, while the effective scale of the derivative coupling (\ref{L_int1}) can be probed up to $10^5$~GeV.
For the small (interior) halo, due to the weaker constraint on the halo mass, the Yukawa-like coupling down to $10^{-21}$ and the effective scale of the derivative coupling up to $10^{10}$~GeV, can potentially be observed.
Constraints can also be obtained for a small halo comprising massive, $U(1)$ vector particles, where the vector-neutrino current coupling as small as $10^{-28}$ can be probed.
These bounds assume a sensitivity to detect an $\sim 10$\% deviation from standard neutrino oscillations. 
If the sensitivity is higher (see, e.g., \cite{Dev:2020kgz}), the bounds are lowered proportionally.

There are several interesting avenues to extend the analysis in this work.
These include performing a more complete analysis of 3-flavour oscillations and studying the effects of CP-violation.
Nevertheless, we expect that the bounds on the dark matter-neutrino couplings discussed in this paper will remain qualitatively the same in the 3-flavour analysis.
Furthermore, it would be interesting to study effects from other neutrino sources, such as the Sun.
Next, our study of the vector halo was restricted to the simple radially-polarised case, hence the similar effects between the vector-neutrino current coupling and the derivative coupling to the scalar halo.
It would be interesting to study other types of polarisation, since this will introduce an additional directional dependence and daily modulations in the neutrino data (see, e.g., \cite{Brzeminski:2022rkf}).
Finally, other possible interactions between the dark matter and Standard Model fields can be considered.
For example, in the case of an axion, couplings to nuclei and electrons can play a role in the formation and structure of the halo, affecting the predictions of the halo mass.

It is intriguing that a local dark matter halo could exist surrounding the Earth. 
Its possible interactions with neutrinos provide a novel way to search for the elusive dark matter particle in neutrino oscillation experiments.

\section*{Acknowledgments}

We thank Zhen Liu and Emin Nugaev for useful discussions.
This work is supported by the Department of Energy Grant No. DE-SC0011842 at the University of Minnesota.

\appendix

\begin{widetext}

\section{}
\label{sec:AppA}
In this Appendix we give the analytic expressions for the $\A_{\theta,m}$ parameters that appear in \eqref{corr1} and \eqref{corr2}. For the dimension-five interaction (\ref{L_int1}) where $g_{ab}$ is Hermitian, we obtain (see also \cite{Huang:2018cwo})
\begin{equation}\label{As11}
    \A_{\theta 1}= 2\Re(g_{12})\cos 2\theta_0-(g_{22}-g_{11})\sin2\theta_0 \;, ~~~~  \A_{m1}=  2\Re(g_{12})\sin 2\theta_0+(g_{22}-g_{11})\cos2\theta_0 \;,
\end{equation}
\begin{equation}\label{As12}
    \A_{\theta 2}=  -2\A_{\theta 1}\A_{m 1}+ 4 \Im^2 (g_{12})\cot 2\theta_0 \;, ~~~~ \A_{m2}= -\A_{m1}^2+4 |g_{12}|^2+(g_{22}-g_{11})^2  \;.
\end{equation}
Instead, in the case of the marginal interaction \eqref{L_int2}, we obtain
\begin{equation}\label{As21}
 \A_{\theta 1}  = \frac{2}{\sum m_\nu}  \biggl( \Re\bigl[ (m_{11}+m_{22})h_{12}^* + m_{12}^*(h_{11}+h_{22}) \bigr] \cos 2\theta_0 - \Re\bigl[ m_{22}h_{22}^*-m_{11}h_{11}^* \bigr] \sin 2\theta_0  \biggr)\,,
\end{equation}
\begin{equation}\label{As22}
 \A_{m1} = \frac{2}{\sum m_\nu} \biggl( \Re\bigl[ (m_{11}+m_{22})h_{12}^*+m_{12}^*(h_{11}+h_{22}) \bigr]\sin 2\theta_0 + \Re\bigl[ m_{22}h_{22}^*-m_{11}h_{11}^* \bigr]\cos 2\theta_0 \biggr) \;,     
\end{equation}
\begin{equation}\label{As23}
    \begin{split}
 &   \A_{\theta 2} = -2\A_{\theta 1}\A_{m 1} + 4 \Im^2\bigl[ m_{12}h_{11}^*-m_{11}h_{12}^*+m_{22}h_{12}^*-m_{12}h_{22}^* \bigr]\cot 2\theta_0 \\
&\qquad+\frac{2\Delta m_0^2}{(\sum m_\nu)^2}\biggl( 2\Re[(h_{11}+h_{22})h_{12}^*]\cos 2\theta_0 +(|h_{11}|^2-|h_{22}|^2)\sin 2\theta_0 \biggr)\,,
\end{split}
\end{equation}
\begin{equation}\label{As24}
    \begin{split}
&   \A_{m2} = -\A_{m1}^2 + \frac{2}{(\sum m_\nu)^2} \left\{\Delta m_0^2\biggl((|h_{22}|^2-|h_{11}|^2 )\cos 2\theta_0 + 2\Re\bigl[ (h_{11}+h_{22})h_{12}^* \bigr]\sin 2\theta_0 \biggr)\right. \\ 
& +\Re\biggl[ (m_{11}h_{11}^*-m_{22}h_{22}^*)^2 -2m_{11}m_{22}^*h_{11}^*h_{22}  +4\left(m_{11}m_{22}h_{12}^{*2}+ m_{11}m_{12}h_{11}^*h_{12}^* + m_{12}^2 h_{11}^* h_{22}^* + m_{12}m_{22}h_{12}^*h_{22}^* \right. \\ 
& \left.\left. + m_{12}m_{11}^* h_{12} h_{22}^* + m_{12}m_{22}^*h_{12}h_{11}^*\right) \biggr]+2|m_{12}|^2(|h_{11}|^2+|h_{22}|^2)+2(|m_{11}|^2+|m_{22}|^2)|h_{12}|^2 +|h_{11}|^2|m_{11}|^2+|h_{22}|^2|m_{22}|^2 \right\}       
    \end{split}
\end{equation}
where $m_{ij}$ are the elements of the neutrino mass matrix in the flavour basis.

\end{widetext}

\bibliography{Refs}

\end{document}